\documentstyle[prd,aps,floats]{revtex}

\begin{document}

\draft
\twocolumn[\hsize\textwidth\columnwidth\hsize\csname
@twocolumnfalse\endcsname

\title{Can topological defects be formed during preheating ?}


\author{S. Kasuya and M. Kawasaki}
\address{Institute for Cosmic Ray Research, University of Tokyo,
  Tanashi, Tokyo 188, Japan}

\date{\today}

\maketitle

\begin{abstract}
    We study the dynamics of a scalar field $\Phi$ with the potential 
    $g(|\Phi|^2-\eta^2)^2/2$ ($g=$ self-coupling constant,
    $\eta=$ symmetry breaking scale) after inflation and make clear
    whether topological defects can ever be formed during
    preheating. In particular, we pay attention to GUT defects
    ($\eta \sim 10^{15}{\rm GeV} - 10^{17}{\rm GeV}$), 
    and consider three types of fluctuations. The first one is
    produced due to parametric resonance, the second is due to the
    negative curvature of the potential, and the last is
    created during inflation. We search for the
    parameter region that nonthermal fluctuations of the scalar field
    produced through the parametric resonant decay of its 
    homogeneous part do not lead to defect formation. We find that
    this region is rather wide, and the GUT defects are not 
    produced after inflation. This fact shows that the
    positiveness of the effective mass square of the field and
    production of large fluctuations whose amplitude is as large as
    that of homogeneous mode are not enough conditions for full
    symmetry restoration. 
\end{abstract}

\pacs{PACS numbers: 98.80.Cq, latex11.27.+d
      \hspace{5cm} hep-ph/9703354}

\vskip2pc]

\setcounter{footnote}{1}
\renewcommand{\thefootnote}{\fnsymbol{footnote}}

\section{Introduction}
The inflation models \cite{Guth,Sato} were invented to solve several
cosmological problems such as the flatness problem and horizon
problem. Among many inflation models it seems that the most natural
and simplest model is chaotic inflation \cite{Chaotic}, in which
one only needs a flat potential for a scalar field called the
inflaton. The quantum fluctuations of the inflaton field during
inflationary epoch become the density fluctuations with
scale-invariant 
spectrum which accounts for the structure formation of the
universe. The amplitudes of the fluctuations are
determined by the self-coupling constant $\lambda$ of the inflaton
field $\phi$ (for the $V=(\lambda/4)\phi^4$ model), which should be
$\simeq 10^{-13}$ \cite{Salopek} to explain the anisotropies of 
Cosmic Microwave Background (CMB) radiation observed by the Cosmic
Background Explorer (COBE) \cite{COBE}. 

In order that the radiation-dominated universe should be recovered
after inflation, the reheating process is necessary for
transferring the vacuum energy to relativistic particles. The inflaton
field that is oscillating at the bottom of its effective potential
decays into other lighter particles due to its coupling to other
particles. In the old version of reheating theory
\cite{old-reheat}, the decay rate is 
estimated by using perturbation theory (Born approximation), and the
reheating temperature can be estimated as  
$T_{{\rm RH}} \simeq 10^{-1}\sqrt{\Gamma_{{\rm tot}}M_p}$
\cite{Linde-book}, based  
on the single particle decay. Here $\Gamma_{{\rm tot}}$ is the total
decay rate and $M_p$ the Planck mass.

Recent investigations revealed that the explosive decay of the
inflaton field takes place in the first stage of reheating 
\cite{KLS1,Shtanov,Boyan1,Yoshi,Kaiser,Tkachev1,KK,Son,AC,PR} 
due to the effect of parametric 
resonance. There are three stages in the reheating process: Inflaton
field decays drastically into bose particles with broad
resonance. This first stage is called {\it preheating}
\cite{KLS1}. After the back reactions of produced particles become
significant, the resonant decay goes into the narrower band. At the
last stage the scatterings and further decays of created particles
occur and the thermal equilibrium is achieved, which is the
completion of the reheating process. 

Particle production can be described by a Mathieu-type
equation or Lam\'{e}-type equation, which has an unstable solution in 
some regions of parameters (instability bands). If the relevant
parameters stay in the instability bands long enough, the solution
grows exponentially. This means that the number of particles produced
becomes exponentially large so that the parametric resonant decay of
the inflaton takes place very efficiently. The physical meaning
of this phenomenon is considered as an {\it induced} effect, in
the sense that the presence of produced particles stimulates the
inflaton field to decay into those particles \cite{KK}. This
phenomenon is thus peculiar to those particles that obey Bose-Einstein
statistics, i.e., bosons. This is the reason that the parametric
resonant decay into fermions cannot occur due to Pauli's exclusion
principle.  

Bosons produced at the preheating stage are far from thermal
equilibrium and have enormously large occupation numbers in the 
low-energy side of their spectrum. One of surprising results of these
nonthermal fluctuations (created particles) during preheating is the
symmetry restoration which may lead to the formation of topological 
defects \cite{KLS2} (see also \cite{Tkachev2}). Assuming that the
inflaton potential is given by  $(\lambda/4)(\phi^2-\eta^2)^2$,
the inflaton $\phi$ decays into $\phi$ particles through parametric
resonance and produced particles result in large fluctuations 
$\langle \delta\phi^2 \rangle \gg \eta^2$, the effective potential of
the inflaton is changed such as it has no negative mass, i.e., the 
symmetry is restored. Then, the restored symmetry is spontaneously
broken again
to produce topological defects. If the symmetry-breaking scale is the
grand unified theory (GUT) scale, GUT defects are produced. The
GUT monopoles or 
domain walls created after inflation destroy the advantage of
introducing the inflationary universe. On the other hand, the GUT
cosmic string may be another seed for the large scale structure
formation. In Ref.\cite{KLS2}, however, the
details of the dynamics of the inflaton field were not studied.   

In this paper, we study the dynamics of both real and complex scalar
field which has a flat potential
$V(\Phi)=\frac{g}{2}(|\Phi|^2-\eta^2)^2$ and is responsible for
topological defects, using numerical calculations including the effect
of the resonant decay and taking into account the back reactions of
fluctuations during preheating in an expanding universe. 
Naively, the topological defects are formed for this
type of potential if there are very large fluctuations of $\Phi$
($|\delta\Phi| \gg |\Phi|$) which implies the full symmetry
restoration. As mentioned above, in Ref.~\cite{KLS2}, it is pointed
out that these fluctuations are produced by the effect of parametric 
resonance. Besides, it will be seen later that the fluctuations also
grow efficiently by the effect of the negative curvature of the
potential of the field $\Phi$. 

However, there is a much more efficient mechanism for producing
topological defects. It is due to the facts that the
field has small fluctuations at the end of inflationary stage and that
the potential of the field has two minima in the radial direction
(After inflation, the phase of the field $\Phi$ is almost fixed  
and it is sufficient to consider only the dynamics of the radial
direction, and the field has small fluctuations beyond the horizon
produced at the inflationary epoch. See below). In this mechanism, the
radial part of $\Phi$ oscillates and settles down to one of two minima
($\pm\eta$) of the potential, and even if there are not so large
fluctuations the final values of $\Phi$ may be different in the
different regions in the universe, which leads to the formation of
topological defects. It is this mechanism that we mainly considered in
this paper. We search for the parameter space where the topological
defects are produced. 

For a real scalar field, we find that the topological
defects are not formed for the GUT scale models 
($\eta \simeq 10^{15} - 10^{17}{\rm GeV}$), even if the effective
mass square of the field 
$m^2_{\Phi,{\rm eff}}$ becomes positive. The crucial point is that 
nonthermal fluctuations cannot be as large as the amplitude of the
coherent mode of the field during preheating. In fact they are smaller 
by two orders of magnitude: 
$\langle \delta\phi^2 \rangle \sim 0.01 \phi^2$. 
Furthermore, even though the amplitude of fluctuations becomes the
same order as that of homogeneous mode due to the effect of 
the negative curvature of the effective potential, the field is
dragged by its classical motion (homogeneous part) so that the field
in the entire universe settles down to one minimum of its potential. 
Therefore, in the case of GUT topological defects, the
field dynamics (the final value of the field $\Phi$) is determined
only by the initial value of its coherent mode and the nonthermal
fluctuations do not affect decisively. On the other hand, however,
if we are allowed to take a breaking scale much smaller than the GUT 
scale, there is formation of topological defects because of the
initial fluctuations due to inflation.

For a complex scalar field, fluctuations in the phase direction grow
much more 
rapidly and the dynamics of homogeneous mode is affected in some
extent. This is because the field $\Phi$ does not feel any potential
in the phase direction. As a result, the amplitude of the fluctuations
are indeed grows as large as that of the homogeneous mode. However, we
find that the topological defects are not formed in the wide region of
the parameter space for the GUT scale models. Therefore, we can
conclude that the defect formation due to the parametric resonance is
very unlikely, even if the effective mass 
square of the field $\Phi$ is positive and the amplitude of
fluctuations becomes large. We also find that topological defects  
are formed if the breaking scale is lower than $10^{14} {\rm GeV}$ such 
as axion models \cite{KKY} because there are enough time for
fluctuations to affect the dynamics of homogeneous part due to, say,
the effects of the narrow resonance.

In Sec. II, we present our models and formalism which are used in
this paper. We study the real scalar field for defects production and
find that the GUT defects cannot be formed in Sec. III. Section IV
is devoted to the study of the complex field. Finally, we make our
conclusions and discussions in Sec. V.  

\section{Model and Formalism}
We consider the dynamics of a scalar field $\Phi$. Since this
field is responsible for the formation of topological defects,
one can write its Lagrangian as
\begin{equation}
    \label{lagrangian}
    {\cal L} = \partial_{\mu}\Phi\partial^{\mu}\Phi^*
               - \frac{g}{2}(|\Phi|^2-\eta^2)^2,
\end{equation}
where $g$ is self-coupling constant and $\eta$ symmetry-breaking 
scale. Here the scalar field $\Phi$ is real or complex, and the produced
topological defects are domain walls or cosmic strings, respectively.
We assume that $g$ is very small, otherwise  the field
$\Phi$ will reach the minimum of its potential very soon during
inflation and no topological defects are formed. (Here we suppose that
the inflaton is another field. See below for more discussions.)
The equation of motion is 
\begin{equation}
  \label{eom-1}
    \partial_{\mu}\partial^{\mu}\Phi + 3H\dot{\Phi}
                            + g(|\Phi|^2-\eta^2)\Phi = 0,
\end{equation}
where $H(=\dot{a}/a)$ is the Hubble constant. When we study the
parametric resonant decay of the complex field $\Phi$, it is more
convenient to 
use two real field $X$ and $Y$ which are defined by $\Phi=X+iY$ with
initial conditions~\footnote{
In this section, formulas are written assuming that $\Phi$ is a
complex field, since the formulas for real $\Phi$ is trivial.}
$X(0)=|\Phi(0)|$ and $Y(0)=0$.
We take the
initial time at the end of inflation. Fluctuations of $X$ and $Y$ can
be represented in terms of the annihilation and creation operators
\begin{eqnarray}
    \delta X(t,\vec{x}) & = & \frac{1}{(2\pi)^{3/2}}\int d^3k 
          (  a_x(k)\delta X_k(t) e^{-i\vec{k}\vec{x}} \nonumber \\
          & & \hspace{30mm}
           + a_x^{\dagger}(k)\delta X_k^*(t) e^{i\vec{k}\vec{x}} ), \\ 
    \delta Y(t,\vec{x}) & = &  \frac{1}{(2\pi)^{3/2}}\int d^3k 
          (  a_y(k)\delta Y_k(t) e^{-i\vec{k}\vec{x}} \nonumber \\
          & & \hspace{30mm}
           + a_y^{\dagger}(k)\delta Y_k^*(t) e^{i\vec{k}\vec{x}} ),
\end{eqnarray}
where $a_i(k)$ and $a_i^{\dagger}(k)$ satisfy commutation relations:
$[a_i(k),a_i^{\dagger}(k^{\prime})]=\delta(\vec{k}-\vec{k}^{\prime})$
with $i=x,y$. If we decompose these fields into a homogeneous mode and
fluctuations, Eq.(\ref{eom-1}) can be rewritten as
\begin{eqnarray}
    & & \ddot{X} + 3H\dot{X} + g(X^2-\eta^2)X \nonumber \\  
           & & \hspace{20mm} = -3g\langle(\delta X)^2\rangle X 
                               - g\langle(\delta Y)^2\rangle X, \\ 
    \lefteqn{\delta\ddot{X}_k + 3H\delta\dot{X}_k 
            + \left[ \frac{k^2}{a^2}-g\eta^2+3gX^2 \right] \delta X_k}
            \nonumber \\ 
    & & \hspace{20mm} = - 3g\langle(\delta X)^2\rangle\delta X_k 
                        -  g\langle(\delta Y)^2\rangle\delta X_k, \\ 
    \lefteqn{\delta\ddot{Y}_k + 3H\delta\dot{Y}_k 
            + \left[ \frac{k^2}{a^2}-g\eta^2+gX^2 \right] \delta Y_k}
            \nonumber \\
    & & \hspace{20mm} = -  g\langle(\delta X)^2\rangle\delta Y_k 
                        - 3g\langle(\delta Y)^2\rangle\delta Y_k, 
\end{eqnarray}
where $\langle \cdots \rangle$ denotes average over space. The 
right-hand side of these equations represent back reactions due to
fluctuations and 
we have used the mean field approximations here (e.g., $\delta X^3
\simeq 3\langle(\delta X)^2\rangle\delta X$,...). This approximation
corresponds to neglecting rescattering of particles~\footnote{
Not all the effect of scatterings are neglected in this
approximations. Actually, forward scatterings are taken into account.}
, but we can
justify it in the following way. Since the typical energy of produced
particles is estimated as $E \sim g^{1/2}|\Phi|$ \cite{KLS1}, the
number density is $n \sim g|\Phi|^4/E \sim g^{1/2}|\Phi|^3$. On the
other hand, the cross section of scattering is approximately given by 
\begin{equation}
    \sigma \sim \frac{g^2}{E^2} \sim \frac{g}{|\Phi|^2}.
\end{equation}
Then the scattering rate becomes
\begin{equation}
    \Gamma \sim n\sigma \sim g^{3/2}|\Phi|.
\end{equation}
What we have to compare with this is the hubble constant 
$H \sim \lambda^{1/2}\phi^2/M_p$:
\begin{equation}
    \frac{\Gamma}{H} 
      \sim g\frac{g^{1/2}|\Phi|}{\lambda^{1/2}\phi}\frac{M_p}{\phi}
      \sim g \frac{M_p}{\phi} \sim g\frac{a(t)}{a(0)},
\end{equation}
where we use $g^{1/2}|\Phi| \sim \lambda^{1/2}\phi$
(see Eq.(\ref{field-evo}) below). Therefore, $\Gamma \ll H$ for
$g(a(t)/a(0)) \ll 1$, which means that the rescattering effect may be
neglected before the time when $a(t)/a(0) \ll g^{-1}$.

Rescaling $t,X$, and $Y$ as 
\begin{eqnarray}
  \label{rescale-1}
    a(\tau)d\tau & = & \sqrt{g}X(0)a(0)dt, \\
  \label{rescale-2}
    x & = & \frac{Xa(\tau)}{X(0)a(0)}, \\
  \label{rescale-3}
    y & = & \frac{Ya(\tau)}{X(0)a(0)},
\end{eqnarray}
we obtain the rescaled dimensionless equations
\begin{eqnarray}
  \label{eom-homo}
    x^{\prime\prime} - \tilde{\eta}^2 a^2 x + x^3 
       + 3\langle\delta x^2\rangle x + \langle\delta y^2\rangle x
       & = & 0, \\  
  \label{eom-x}
    \delta x^{\prime\prime}_k 
       + [ \tilde{k}^2 - \tilde{\eta}^2 a^2 + 3x^2
          + 3\langle\delta x^2\rangle  
          +  \langle\delta y^2\rangle ] \delta x_k & = & 0, \\
  \label{eom-y}
    \delta y^{\prime\prime}_k 
       + [ \tilde{k}^2 - \tilde{\eta}^2 a^2 +  x^2
          +  \langle\delta x^2\rangle  
          + 3\langle\delta y^2\rangle ] \delta y_k & = & 0,    
\end{eqnarray}
where we define $\tilde{k}=k/\sqrt{g}X(0)$, $\tilde{\eta}=\eta/X(0)$,
and $a(0)=1$. The prime denotes differentiation with respect to $\tau$
and we assume that the universe is radiation dominated, which is 
a reasonable approximation.

If the universe is radiation dominated, the Hubble parameter evolves
as 
\begin{equation}
  \label{hubble}
    H(\tau) = H(0) \left( \frac{a(0)}{a(\tau)} \right)^2.
\end{equation}
Since the potential energy of the inflaton field dominates
the universe during inflation, the Hubble parameter at the end of
inflation is
\begin{equation}
  \label{hubble-0}
    H(0) = \sqrt{\frac{2\pi}{3}}\frac{\sqrt{\lambda}}{M_p}\phi^2(0), 
\end{equation}
where we assume that the inflaton potential is
$V(\phi)=(\lambda/4)\phi^4$ ($\lambda=10^{-13}$). First let us
consider the case that the field $\Phi$ which we are concerned with is
the inflaton, i.e., $\Phi = \phi$ (or $g=\lambda$). From
Eqs.(\ref{hubble}) and (\ref{rescale-1})  
\begin{equation}
    H(\tau) =
       \sqrt{\lambda}\phi(0)a(0)\frac{a^{\prime}(\tau)}{a^2(\tau)}  
            = H(0) \left( \frac{a(0)}{a(\tau)} \right)^2.
\end{equation}
Integrating this equation and use Eq.(\ref{hubble-0}), we get
\begin{equation}
    a(\tau) = \sqrt{\frac{2\pi}{3}}\frac{\phi(0)}{M_p}a(0)\tau + a(0).     
\end{equation}
As the inflationary epoch ends at the time when $\phi=M_p/\sqrt{3\pi}$ 
\cite{Linde-book}, and $a(0)=1$ is imposed, the evolution of the
scale factor becomes
\begin{equation}
  \label{a-evolve}
    a(\tau) = \frac{\sqrt{2}}{3}\tau + 1.
\end{equation}
Next we consider the case that the inflaton field is different from
$\Phi$ field. Then the Hubble parameter becomes
\begin{equation}
    H(\tau) = \sqrt{g}X(0)a(0)\frac{a^{\prime}(\tau)}{a^2(\tau)} 
                  = H(0) \left( \frac{a(0)}{a(\tau)} \right)^2.
\end{equation}
Integrating this equation, we obtain
\begin{equation}
  \label{a-other}
    a(\tau) = \frac{\sqrt{2\pi}}{3}\frac{\phi(0)}{M_p}
              \frac{\sqrt{\lambda}\phi(0)}{\sqrt{g}X(0)}\tau + 1,
\end{equation}
where $a(0)=1$ is assumed again. Since the classical (homogeneous)
mode $X$ slowly evolves as 
\begin{equation}
  \label{field-evo}
    X \simeq \left( \frac{\lambda}{g} \right)^{1/2} \phi
\end{equation}
in the inflationary epoch, Eq.(\ref{a-other}) becomes identical to
Eq.(\ref{a-evolve}) using $\phi(0)=M_p/\sqrt{3\pi}$ again. Thus we can 
use Eq.(\ref{a-other}) for a general scalar field $\Phi$.

Now let us consider the initial conditions. The initial condition
for the homogeneous mode is $x(0)=1$ from its definition. 
The value of the field $\Phi$ is 
initially almost constant in the whole universe because of the
inflation. However, during inflation, the field has small quantum
fluctuations, which become seeds for the large scale structures of
the universe. We take account of these small fluctuations in the 
following way. In the inflationary epoch, the scalar field fluctuates
with the amplitude $H/2\pi(\sim 10^{-6}(g/\lambda)^{1/2}X(0))$. These
fluctuations are stretched beyond the horizon size by inflation and
become classical. Therefore,
one can regard that the initial value will be
slightly changed as  $X(0) \rightarrow \tilde{X}(0)=X(0)(1+\Delta)$,
where $\Delta \sim 10^{-6}\times (g/\lambda)^{1/2}$, in each region of 
the universe~\footnote{
Here the relevant size of the region is between 
$\sim H^{-1}(\eta/M_p)^{-1}$ (which corresponds to the horizon at the
defect formation) and $\sim H^{-1}e^{60}$ (which corresponds to the
present horizon) when $|\Phi| \simeq \eta$.}, 
and $X(0)$ is considered as the mean initial value. Then
the initial condition for homogeneous mode must be $x(0)=1+\Delta$.  

Next we consider the initial values of fluctuations. 
Hamiltonian is a diagonalized operator for a free field, but in the 
presence of interactions, it is not diagonalized in terms of $a$ and
$a^{\dagger}$. It can be diagonalized at any instant of time by means
of Bogoliubov transformations, which relate annihilation and
creation operators $a$ and $a^{\dagger}$ at $t=0$ to time dependent
annihilation and creation operators $b(t)$ and $b^{\dagger}(t)$: 
\begin{eqnarray}
    b_X(t) & = & \alpha_X(t)a_X+\beta_X^*(t)a_X^{\dagger}, \\ 
    b^{\dagger}_X(t) & = & \beta_X(t)a_X+\alpha_X^*(t)a_X^{\dagger}, 
\end{eqnarray}
and $\alpha_X(t)$ and $\beta_X(t)$ can be written as
\cite{Shtanov}
\begin{eqnarray}
    \alpha_X & = & \frac{e^{ i\int\Omega_Xdt}}{\sqrt{2\Omega_X}}
                        (\Omega_X\delta X_k + i\delta\dot{X}_k), \\
    \beta_X  & = & \frac{e^{-i\int\Omega_Xdt}}{\sqrt{2\Omega_X}}
                        (\Omega_X\delta X_k - i\delta\dot{X}_k),
\end{eqnarray}
where $\Omega_X^2 = k^2/a^2-g\eta^2a^2+3g\tilde{X}^2$, and the
initial conditions for $\alpha_X$ and $\beta_X$ are
\begin{equation}
    |\alpha_X(0)|=1, \hspace{15mm} \beta_X(0) =0,
\end{equation}
which means that there is no particles at $t=0$.
These conditions correspond to 
\begin{equation}
  \label{init-xx}
    |\delta X_k(0)|=\frac{1}{\sqrt{2\Omega_X(0)}}, \quad
    i\delta \dot{X}_k(0)=\Omega_X(0)\delta X_k(0). 
\end{equation}
Similarly, initial conditions for $\delta Y_{\ k}$ are
\begin{equation}
  \label{init-yy}
    |\delta Y_k(0)|=\frac{1}{\sqrt{2\Omega_Y(0)}}, \quad
    i\delta \dot{Y}_k(0)=\Omega_Y(0)\delta Y_k(0), 
\end{equation}
where $\Omega_Y^2 = k^2/a^2-g\eta^2a^2+g\tilde{X}^2$.

Then if we rescale as Eqs.(\ref{rescale-1})-(\ref{rescale-3}), we get 
\begin{eqnarray}
  \label{init-x}
     & & |\delta x_k(0)|=\frac{1}{\sqrt{2\Omega_X(0)}\tilde{X}(0)},  
    \nonumber \\  
     & & \delta x_k^{\prime}(0)
       = \left[ h(0)-i\frac{\Omega_X(0)}{\sqrt{g}\tilde{X}(0)} \right] 
                                                 \delta x_k(0), \\
  \label{init-y}
     & & |\delta y_k(0)|=\frac{1}{\sqrt{2\Omega_Y(0)}\tilde{X}(0)}, 
    \nonumber \\
     & & \delta y_k^{\prime}(0)
       = \left[ h(0)-i\frac{\Omega_Y(0)}{\sqrt{g}\tilde{X}(0)} \right] 
                                                 \delta y_k(0), 
\end{eqnarray}
where $h(\tau)=a^{\prime}(\tau)/a(\tau)$ (from Eqs.(\ref{a-evolve}),
$h(0)=\sqrt{2}/3$), and the average of the square of the fluctuations
can be written as~\footnote{
Equations (\ref{fl-x}) and (\ref{fl-y}) should be renormalized
\cite{Ren}. To this end,  
the vacuum fluctuations are subtracted from these expressions when
calculated in the numerical calculations.} 
\begin{eqnarray}
  \label{fl-x}
    \langle \delta x^2 \rangle 
    & = & \int \frac{d^3k}{(2\pi)^3} |\delta x_k|^2, \\
  \label{fl-y}
    \langle \delta y^2 \rangle 
    & = & \int \frac{d^3k}{(2\pi)^3} |\delta y_k|^2.
\end{eqnarray}

In the following sections, we integrate
Eqs.(\ref{eom-homo})-(\ref{eom-y}) using relation (\ref{a-evolve})
with initial conditions (\ref{init-x}),  
(\ref{init-y}) and $x(0)=1+\Delta$ to see whether topological defects 
are formed or not.
We take enough resolution of momentum space to calculate
Eqs.(\ref{fl-x}) and (\ref{fl-y}) in all cases which we considered.  
 
\section{The real scalar field}
\subsection{The classical dynamics of the field: Only with cosmic
expansion}
First we consider the evolution of the scalar field without 
fluctuations after inflation. We expect that 
the dynamics of the field should be determined entirely
classically. As mentioned above, however, the scalar
field fluctuates with the amplitude $H/2\pi$ in the inflationary
epoch, and the fluctuation will be  
$|\delta\Phi/\Phi| \simeq 10^{-6}\times (g/\lambda)^{1/2}$ at the end
of inflation, which leads that the field in different space points
might fall into the different minima of its effective
potential. Therefore, we should study the dependence of the field
dynamics (the final value of the field) on its initial value. If there
are no dependence on those 
initial values whose differences are as large as $10^{-6}\times
(g/\lambda)^{1/2}$, we can conclude that the evolution of the field
is completely  determined by classical dynamics. Otherwise, we cannot
discuss its classical dynamics, to say nothing of the effects of
nonthermal fluctuations produced in the preheating epoch which will
be considered later. 

 From Eq.(\ref{eom-homo}), the classical equation of motion
without the effect of fluctuations is written as
\begin{equation}
    x^{\prime\prime} - \tilde{\eta}^2 a^2 x + x^3 = 0, 
\end{equation}
where the initial condition is $x(0)=1+\Delta$.

Figure~\ref{fig-homo4} is an example for the time evolution of the field
with $\Delta=0$ (Here we take $g=\lambda=10^{-13}$ which corresponds
to the case that the field $\Phi$ is the inflaton, and $\eta=10^{15} 
{\rm GeV}$). We see that the amplitude of the field $x$ grows larger
than the initial value. This is because we rescaled the field as
Eq.(\ref{rescale-1}) and the amplitude grows as
$\sqrt{1+\tilde{\eta}^2a^2(\tau)}$.

Figure~\ref{fig-scaling} shows the numerical result of the critical
value of $\Delta$ with $\eta=10^{12}-10^{17} {\rm GeV}$ for
$g=\lambda=10^{-13}$. If $\Delta \gtrsim \Delta_{{\rm crit}}$, the
initial 
fluctuations do affect the dynamics of the field (the final value of
the field), which results in the formation of topological defects. 
On the other hand, if $\Delta \lesssim \Delta_{{\rm crit}}$, the
fluctuation 
of the field never affects the final value of the field. 
$\Delta_{{\rm inf}}$ (the amplitude of fluctuations at the end of
inflation) is less than $\Delta_{{\rm crit}}$ for 
$\eta \gtrsim 10^{13} {\rm GeV}$. 
Therefore, the GUT scale topological defects are not produced after
inflation if we neglect the fluctuations of $x$. 

Let us consider the evolution of the scalar field $\Phi$ for
$g > \lambda$ (i.e., $\Phi$ is not an inflaton). As $g$ grows large, 
$\Delta_{{\rm inf}}$ grows as $\Delta_{{\rm inf}}\propto g^{1/2}$. On  
the other hand, the potential of the field can be rewritten as 
$V=\frac{1}{4}[(g^{1/2}\Phi)^2 - (g^{1/2}\eta)^2]^2$. Since we have
relation, $g^{1/2}\Phi(0) \simeq \lambda^{1/2}\phi(0)$
(cf. eq.(\ref{field-evo})), the evolution of the field is equivalent
to that for $g=10^{-13}$ when $\eta$ is
larger by the factor $(g/\lambda)^{1/2}$. Therefore, from our
calculations (Fig.~\ref{fig-scaling}), we find that no topological
defect is formed for $\eta \gtrsim 10^{13} {\rm GeV}$, which is
independent of $g$. 

We can also obtain this result analytically by comparing the initial
fluctuation $|\delta\Phi/\Phi| (\sim 10^{-6} \times
(g/\lambda)^{1/2})$ with the change of the amplitude $\Delta A$ of 
oscillating $\Phi$ due to cosmic expansion in one oscillation time
$T$. Since $\dot{\Phi} \simeq -\frac{3}{2}H\Phi$ at the critical epoch
(when $A \sim \eta$), $\Delta A$ is given by
\begin{equation}
    \frac{\Delta A}{A} 
      \sim HT 
      \sim \frac{\lambda^{1/2}\phi^2}{M_p}(g^{1/2}\eta)^{-1}
      \sim \frac{\eta}{M_p} \left( \frac{g}{\lambda} \right)^{1/2}
      \equiv \Delta_{{\rm crit}},
\end{equation}
where we use eq.(\ref{field-evo}). If $\Delta A/A$ is larger than
$|\delta\Phi/\Phi|$, the initial fluctuations do not affect the
dynamics of the classical evolution of the field $\Phi$. 
Since $\Delta_{{\rm inf}}\equiv |\delta\Phi/\Phi|
\sim 10^{-6} \times (g/\lambda)^{1/2}$, we can  
obtain the condition that the field $\Phi$ settles down to the
definite minimum of the potential in the entire universe:
\begin{equation}
  \label{delta-crit}
    \frac{\eta}{M_p} \gtrsim 10^{-6}. 
\end{equation}
This is the same as what we obtained from our numerical calculations.
Notice that this condition is, of course, independent of the coupling
constant $g$.

\subsection{Preheating stage: Effect of nonthermal fluctuations}
In order to study the production of nonthermal fluctuations and find 
whether the formation of topological defects occurs or not, we must
investigate how fluctuations can be 
created. In fact, they are produced due to both 
parametric resonance during preheating and the negative curvature of
the effective potential. To investigate how
large the non-thermal fluctuations grow, we consider these two
effects separately. First, we study the parametric resonance in
this subsection. The effect of the negative curvature of the potential
will be considered in the next subsection. 

We integrate those equations which the breaking terms are omitted
from Eqs. (\ref{eom-homo}) and (\ref{eom-x}), i.e., $\eta=0$. For the 
early stage (during preheating), neglecting the breaking terms is
quite a reasonable approximation, since the breaking scale is much
smaller than the initial amplitude of the field $\Phi$.
The result is shown in Fig.~\ref{fig-r-long}. The solid line denotes
the envelope of $x^2$ and the dotted line $\delta x^2$: fluctuations
of $x$, and these lines are extrapolated to larger $a$. Note that the  
decaying power of homogeneous mode is $\sim -0.26$, which is smaller
than the value $-1/3$ estimated in the Ref.~\cite{KLS2}.~\footnote{
The value $-1/3$ can be estimated with assumptions that 
$\langle \delta x^2 \rangle$ grows large and dominates the effective
mass square of the field and becomes almost constant. Using this naive
estimation, we get the same decaying power for the both cases that
the evolution of homogeneous mode is written in terms of the
elliptic function and the trigonometric function. However, the
dynamics of fluctuation obeys Lam\'{e} equation, not Mathieu equation,
so that the instability band is narrower and the growth rate is much
smaller in the case of Lam\'{e} equation. Therefore, the decaying
power is smaller.}
This is because the actual dynamics is much
more complicated since $\langle \delta x^2 \rangle$ is also
oscillating for example. (Our value $-0.26$ is much smaller than the
value $\sim -2/3$ in Ref.~\cite{KhTk}. In this reference, however,
the calculation has been done for only a short time such as $\tau \sim 
800$, so that the decaying power in the narrow resonance stage cannot
be seen from such a short time evolution.) The time that the
field falls into one minimum of its potential is estimated as 
$a_* \sim (\eta/X(0))^{-1}$. For the GUT scale ($\eta \sim 10^{16}
{\rm GeV}$), $a_* \sim 400$. At
this time the amplitude of fluctuations is much less than that of
homogeneous mode. Therefore, the fluctuations is too small
to produce the GUT defects. On the other hand, we can see that the
crossing takes place at the time $a_c \sim 5 \times 10^5$. For
$a>a_c$, the dynamics of the field $\Phi$ may be fully affected
by its fluctuations since the field fluctuates more than 
${\cal O}(1)$, which means the topological defects may be developed in 
the same sense as pointed out in Ref.~\cite{KLS2}.
Therefore, IF $\eta \lesssim 10^{13} {\rm GeV}$,
it is expected that the full symmetry 
restoration occurs and leads to the formation of
topological defects due to the later spontaneous symmetry breaking (if 
there is any breaking term in the equation). 

\subsection{Postpreheating: Effects of the negative curvature of the
potential}
Now let us take into account the effects of the negative curvature of 
the potential. To this end, we integrate equations
\begin{eqnarray}
  \label{x-real-eom}
    x^{\prime\prime} - \tilde{\eta}^2 a^2 x + x^3  
                     + 3\langle\delta x^2\rangle x & = & 0, \\
  \label{dx-real-eom}
    \delta x^{\prime\prime}_k 
       + [\tilde{k}^2 - \tilde{\eta}^2 a^2 + 3x^2 
                     + 3\langle\delta x^2\rangle ] \delta x_k & = & 0,
\end{eqnarray}
and see how large the amplitude of the fluctuations grows 
comparing with that of the homogeneous mode. The time evolution of the 
homogeneous mode and the fluctuations are shown in
Figs.~\ref{fig-real1} and ~\ref{fig-real-fl}, respectively. From
these figures we see that the preheating stage ends at $a \sim
250$. What we must 
pay attention to is the fact that the amplitude of nonthermal
fluctuations during preheating is two orders smaller than that of
homogeneous mode, which is contrary to the estimation of
Ref.~\cite{KLS2} (It can also be seen from Fig.~\ref{fig-r-long}, and
is consistent with Ref.~\cite{KhTk}). Moreover, 
the amplitude grows due to the global instability of negative mass
(breaking term), but the maximum amplitude is at  
most a few factor less than that of the homogeneous mode (at $a \sim
3000$). This is the crucial point. We can say in other words as
follows: although the effective mass square of the field becomes
positive during preheating (see Fig.~\ref{fig-real-mphi}), the amplitude
of the fluctuations is smaller than that of homogeneous one so that no
full symmetry restoration takes place, which results in no development
of GUT topological defects (domain walls). Notice that the typical
momentum of the fluctuations which
contributes to fluctuations $\langle \delta x^2 \rangle$ is larger
than $H \sim \sqrt{\lambda}\phi \sim \sqrt{g}|\Phi|$ (see
Fig.~\ref{fig-real-sp}). This means that it is a good approximation to
take the conformal vacuum for initial value of fluctuations.

\subsection{Dynamics of the field with the initial fluctuations} 
As we mentioned in Sec. II, there is much more efficient mechanism
for producing topological defects. It is due to the facts that the
field has small fluctuations at the end of inflationary stage and that
the potential of the field has two minima in the radial direction. In
this mechanism, the radial part of $\Phi$ oscillates and settles down
to one of two minima ($\pm\eta$) of the potential, and even if there
are not so large fluctuations the final values of $\Phi$ may be
different in the different regions in the universe, which leads to the
formation of topological defects. In order to study this effect is the
main investigation of this paper, and we show it in this subsection.

We consider the dynamics of the field taking into account of
the initial fluctuations created during inflation in the following
way: As we discussed in Sec. II, we numerically integrate
Eqs. (\ref{x-real-eom}) and (\ref{dx-real-eom}) with initial 
conditions (\ref{init-x}) for fluctuations and 
$x(0)=1+\Delta$ for a homogeneous mode, where $\Delta$ denotes the
amplitude of the initial fluctuations, 
and see how much the dynamics of the homogeneous mode is affected by
those nonthermal fluctuations produced both during preheating due to
the parametric resonance and due to the negative
curvature of the potential. 

We can see little effects of the nonthermal fluctuations, since the
difference between Figs.~\ref{fig-real-scale} and ~\ref{fig-scaling} 
cannot be seen. Therefore, we conclude that
topological defects are not formed for
$\eta \gtrsim 10^{13} {\rm GeV}$ and any value of $g$, which means
that the GUT defects are not be produced during reheating after
inflation. This is the same conclusion as the case without considering 
the nonthermal fluctuations.

\section{The complex scalar field}
\subsection{Preheating stage: Effect of nonthermal fluctuations}
Let us turn our attention to the dynamics of a complex field
$\Phi=X+iY$. As mentioned previously, there are three kind of
fluctuations: the nonthermal fluctuations created due to parametric
resonance, the one due to the effect of the negative curvature of
the potential, and the initial fluctuations produced during inflation. 
In this subsection, we pay attention to the first one. To this
end, omitting the breaking term, we have integrated the following
rescaled equations: 
\begin{eqnarray}
    x^{\prime\prime} + x^3 + 3\langle\delta x^2\rangle x 
                           +  \langle\delta y^2\rangle x & = & 0, \\  
    \delta x^{\prime\prime}_k 
     + [ \tilde{k}^2 + 3x^2 
        + 3\langle\delta x^2\rangle  
        +  \langle\delta y^2\rangle ] \delta x_k & = & 0,\\ 
    \delta y^{\prime\prime}_k 
     + [ \tilde{k}^2 +  x^2
        +  \langle\delta x^2\rangle  
        + 3\langle\delta y^2\rangle ] \delta y_k & = & 0.
\end{eqnarray}
The result is shown in Fig.~\ref{fig-c-long}. The solid line denotes
the envelope of $\varphi^2=x^2$ and the dotted line fluctuations,
i.e., $\langle \delta\varphi^2 \rangle = \langle \delta x^2 \rangle +
\langle \delta y^2 \rangle$, which are extrapolated
to larger $a$. At $a_* (\sim (\eta/X(0))^{-1}) \sim 400$, there are
not large enough fluctuations to produce the GUT defects. 
We see that the crossing time is $a_c \sim 10^5$. This
means that topological defects may be formed for the models with
breaking scale $\eta \lesssim 3\times 10^{13} {\rm GeV}$. 
Note that the amplitude of fluctuations is larger than that
of the real scalar field. This is because the phase fluctuation
$\delta \theta$ (Goldstone mode) approximately written as $\delta y/x$
feels no potential so that it can grow much faster \cite{Boyan1,KKY}.
Therefore, the crossing time is earlier than the real scalar field
case even if the decaying power ($ \simeq -0.185$) is smaller due to
large back reaction from the phase fluctuation.

\subsection{Postpreheating: Effects of the negative curvature of the
potential}
Now we must include the breaking term in the equations in order to see 
the effect of the negative curvature of the potential on producing
fluctuations. We integrate Eqs.(\ref{eom-homo}) - (\ref{eom-y})
numerically. The dynamical evolution of the homogeneous mode with taking
account of the effects of non-thermal fluctuations is shown in
Fig.~\ref{fig-com3}. The field decays abruptly at 
$a \sim 50$. This time is earlier than 
that for the case of the real scalar field. The reason can be seen in 
Fig.~\ref{fig-com-fl-Y} (Figs.~\ref{fig-com-fl} and
~\ref{fig-com-fl-Y} are the evolutions of radial and phase
fluctuations, respectively, when $\Delta=0$). It is reasonable that the
amplitude of $\delta y$ grows much faster than that of $\delta x$
(compare Fig.~\ref{fig-com-fl} with Fig.~\ref{fig-com-fl-Y}),
since the fluctuation $\delta y$ has an approximate relation 
$\delta \theta \simeq \delta y/x$ to the phase fluctuation which can
be identified with a Goldstone mode who feels no potential, as
previously mentioned. Therefore, the energy stored in the homogeneous
mode of the field was quickly transferred into this Goldstone mode
first. This is also seen in the momentum distribution of $\delta x$
and $\delta y$ (Figs.~\ref{fig-com-sp} and ~\ref{fig-com-spY}).   

We see, however, that nonthermal fluctuations are not so much
produced in the preheating epoch. At the end of preheating, the
amplitudes of nonthermal fluctuations are $\langle (\delta x)^2
\rangle \sim  0.01x^2, \langle (\delta y)^2 \rangle \sim 0.1 x^2$,
which are smaller than those 
estimated ($\sim x^2$) in the Ref.~\cite{KLS2}. Therefore,
topological defects cannot be directly formed due to the effects of
nonthermal fluctuations produced, even if the effective mass square
of the field is positive during preheating (see
Fig.~\ref{fig-com-mphi}). Even though the amplitude of fluctuations
becomes as large as that of the homogeneous mode at $a \sim 3000$ due
to the effect of the negative curvature of the potential, it
is still not enough condition for the formation of topological
defects (It will be verified in the next subsection). 

\subsection{Dynamics of the field with the initial fluctuations}
As mentioned many times above, there are fluctuations produced at
the inflationary epoch. Therefore, the initial value of the field
$\Phi$ is slightly different in each region which contains many
horizons, and we take the initial 
condition as $x(0)=1+\Delta$. In order to consider the effects of
nonthermal fluctuations to the dynamics of the homogeneous mode in
the presence of this initial fluctuations, we 
integrate Eqs.(\ref{eom-homo}) - (\ref{eom-y}) 
with initial conditions (\ref{init-x}) and (\ref{init-y}) for
fluctuations and $x(0)=1+\Delta$ for the homogeneous mode.

Figure~\ref{fig-complex-scale1} is the result for $g=10^{-13}$. The
evolution of the homogeneous mode (the final value of the field) is
affected if the breaking scale $\eta \lesssim 10^{14} {\rm GeV}$.
Comparing with the results without including fluctuations (the
dashed line), the fluctuations considerably affect the dynamics of the 
homogeneous field (the final value of the field), and it is different
from the real scalar field case. This is because the large phase
fluctuations are produced.
There are some unstable region about $\eta \simeq 10^{16} {\rm GeV}$. 
For $\eta \sim 10^{16} {\rm GeV}$, the critical time
when the field settles down to one of the minimum of the potential is
almost the same time when the radial fluctuations become
significant. We think that this is the reason for the unstable
behavior of $\Delta$ around $\eta = 10^{16} {\rm GeV}$.

For $g>\lambda$, we can study the dynamics of $\Phi$ by rescaling
$\Delta_{{\rm inf}}$ and $\eta$ from the result for
$g=\lambda=10^{-13}$. In Fig.~\ref{fig-complex-scale2} we show the
region of the defects formation in the $g - \eta$ plane. From this
figure it is seen that the topological defects are not formed for
almost all the region of the GUT scale 
($\eta = 10^{15} - 10^{17} {\rm GeV}$). 

Nonthermal fluctuations do actually affect the evolution 
when comparing with the results for the classical evolution
without fluctuations studied in the previous section. Owing to the
effects of nonthermal fluctuations, the region where topological
defects are produced are larger than that for the pure classical
one. Note that the GUT symmetry is not 
fully restored during preheating in spite of the fact that the
effective mass square of the field becomes positive before the
field settles down to the minimum of its effective potential. This is 
because that the amplitude of fluctuations do not highly overcome that
of the homogeneous mode, and the fluctuations cannot disturb the 
whole dynamics of the field even if its amplitude becomes as large as
that of homogeneous mode. The authors in Ref.\cite{KLS2} assert
that the criterion of symmetry restoration is the sign of the
effective mass square of the field. As we mentioned above, this
condition is not enough to judge the symmetry restoration. They
consider that the amplitude of fluctuations is as large as that of
homogeneous mode at the end of preheating and the former always
overcome the latter due to the narrow resonance effects in the
following stage. The very difference from their conclusion is that
the fluctuations of the filed $\Phi$ are not created so much and hence
there needs much longer period of narrow resonance stage for the
amplitude of fluctuations to be much larger than that of homogeneous
one. 

Another viewpoint is insisted by the authors of
Ref.\cite{Boyan2}: the symmetry is always restored if there 
is a very large energy density in the initial state enough for the
field to occupy both vacua under the dynamical evolution. However, we
see that the topological defects will not be formed even if initial
energy is large. Therefore, the criterion given in Ref.~\cite{Boyan2}
is also not enough condition.

\section{Conclusions and Discussions}
We consider the dynamics of both the real and complex scalar field
$\Phi$ with potential $(g/2)(|\Phi|^2-\eta^2)^2$ after inflation. In
the preheating phase, the homogeneous mode of the field $\Phi$
transfers its energy into nonthermal fluctuations through the
parametric resonance, and their amplitudes are expected much larger
than that in 
the thermal equilibrium. These nonthermal fluctuations may cause the
restoration of symmetry and hence the formation of topological
defects. In the present work, we consider tree types of
fluctuations. The first one is produced due to parametric resonance
effects, the second is due to the effect of the negative curvature of
the potential, and the last is created during inflation. 

To investigate these fluctuations thoroughly, we numerically
integrated the coupled equations for a homogeneous mode and
fluctuations using the mean field approximations for the very flat
potential $V(\Phi)=(g/2)(|\Phi|^2-\eta^2)^2$. At the
inflationary epoch, the field $\Phi$ fluctuates and its
amplitude becomes $\delta\Phi=H/2\pi$. Therefore, at the end of
inflation, $\delta\Phi/\Phi \sim 10^{-6}\times(g/\lambda)^{1/2}$. This 
initial fluctuation might affect the dynamics of the field (the final
value of the field, i.e., $\Phi = \eta$ or $-\eta$) in the reheating
epoch after inflation. On account of the initial fluctuations, we take
the initial value of the homogeneous mode such as $x(0)=1+\Delta$,
which is equivalent to considering each domain of the universe beyond
the horizon with different initial conditions.

First we study the evolution of a classical field with various initial
conditions $\Delta$ when there is no nonthermal fluctuations produced
by the effect of parametric resonance, and find that the
topological defects are formed for $\eta \lesssim 10^{13}  
{\rm GeV}$. The physical meaning of this is very clear. If the
amplitude of initial fluctuations is smaller than the amplitude change
in one oscillation time due to cosmic expansion, it has no influence 
over the field dynamics. Since the energy change can be estimated as 
$\Delta A/A \sim H/\omega_{{\rm osc}} \sim M_{{\rm GUT}}/M_p 
\sim 10^{-4}$ for the
GUT models, $\delta\Phi/\Phi \ll \Delta A/A$ so that the initial
fluctuations cannot affect.

Next we study the dynamics of the real scalar field taking into 
account the effects of nonthermal fluctuations produced as a result
of both the parametric resonance and the effect of the negative 
curvature of the potential in the preheating epoch, and how they
affect the dynamics of the field. We find that fluctuations do not
influence the dynamics of the homogeneous mode at all, and it is
entirely the same as the case of the classical evolution without
fluctuations. Therefore, no GUT topological defects ($\eta \sim
10^{16} {\rm GeV}$) can be formed
even if the amplitude of the fluctuations produced during the
reheating stage grows as large as that of the homogeneous mode. This
means that the symmetry restoration does not occurred in spite of the 
positiveness of the effective mass square of the field $\Phi$.

For the complex scalar field, fluctuations do
affect the evolution of the field, since the phase fluctuation is
produced much more rapidly than that in the radial direction since
the field feel no potential in phase direction (Goldstone mode). 
However, there is no but a tiny region for the formation of
topological defects for the GUT scale in the parameter space, even if 
fluctuations are much produced and the effective mass square of the
field becomes positive during preheating. In other words, it is not
enough condition for the symmetry restoration that the field has a
positive effective mass square and fluctuates at most of order 
${\cal O}(1)$. On the other hand, topological defects are formed 
in the model with much lower breaking scale ($\eta \lesssim 10^{14}
{\rm GeV}$) such as axion models.

\section*{Acknowledgment}
The authors are grateful to Masahide Yamaguchi and Tsutomu Yanagida
for very illuminating comments and discussions.

\newpage



\begin{figure}
    \caption{ Evolution of the scalar field
    without nonthermal fluctuations (only classical evolution) for 
    the initial deviation $\Delta=0$ and $g=10^{-13}$.
  \label{fig-homo4} }
\end{figure}

\begin{figure}
    \caption{ Critical value of $\Delta$ as a function of
    $\eta$ for $g=10^{-13}$. The stars are from numerical
    calculations. The dashed line shows the analytic estimation of 
    Eq.(\ref{delta-crit}) including some numerical factors, and the
    dotted line denotes the amplitude of the initial fluctuations
    produced in the inflationary epoch. If the latter line is above
    the former one, the dynamics of the field is considerably
    affected, which results in the formation of defects.
  \label{fig-scaling} } 
\end{figure}


\begin{figure}
    \caption{ Long time evolution of the dynamics of the real
    scalar field with nonthermal fluctuations. The dots and stars
    denote the amplitude square of homogeneous part ($x^2$) and
    fluctuations ($\langle \delta x^2 \rangle$), 
    respectively. The solid line stands for the envelope of $x^2$
    and is extrapolated to the large $a$. Similarly, the dotted
    line represents the envelope of $\langle \delta x^2 \rangle$.
  \label{fig-r-long} }
\end{figure}

\begin{figure}
    \caption{ Evolution of the homogeneous mode of the real
    scalar field with nonthermal fluctuations 
    with initial deviation $\Delta=0$. We take 
    $\eta=10^{15} {\rm GeV}$ and $g=10^{-13}$. 
  \label{fig-real1} }
\end{figure}

\begin{figure}
    \caption{ Evolution of the fluctuation of the real scalar
    field 
    with initial deviation $\Delta=0$.
  \label{fig-real-fl} } 
\end{figure}

\begin{figure}
    \caption{ Evolution of the effective mass square at $\Phi=0$
    for the real scalar field 
    with initial deviation $\Delta=0$.
  \label{fig-real-mphi} } 
\end{figure}

\begin{figure}
    \caption{ Power spectrum of the radial fluctuation 
    $\delta x_k$ for (a) $a \simeq 1000$, (b) $a \simeq 2000$, 
    (c) $a \simeq 3000$ and (d) $a \simeq 3500$. 
  \label{fig-real-sp} } 
\end{figure}

\begin{figure}
    \caption{ Critical value of $\Delta$ as a function of
    $\eta$. The stars are from numerical calculations. 
    The dashed line shows the analytic estimation without fluctuations 
    (see Fig.~\ref{fig-scaling}),
    and the dotted line denotes the amplitude of the initial
    fluctuations produced in the inflationary epoch. 
  \label{fig-real-scale} } 
\end{figure}


\begin{figure}
    \caption{ Long time evolution of the dynamics of the complex
    scalar field with nonthermal fluctuations. The dots and stars
    denote the amplitude square of homogeneous part ($\phi^2$) and
    fluctuations ($\langle \delta \phi^2 \rangle$), 
    respectively. The solid line stands for the envelope of $\phi^2$
    and is extrapolated to the large $a$. Similarly, the dotted
    line represents the envelope of $\langle \delta \phi^2 \rangle$.
  \label{fig-c-long} }
\end{figure}

\begin{figure}
    \caption{ Evolution of the homogeneous mode of the complex
    scalar field with nonthermal fluctuations 
    for initial deviation $\Delta=0$ and $g=10^{-13}$.
  \label{fig-com3} }
\end{figure}

\begin{figure}
    \caption{ Evolution of the radial fluctuation $\delta x$ of
    the complex scalar field 
    with initial deviation $\Delta=0$.
  \label{fig-com-fl} } 
\end{figure}

\begin{figure}
    \caption{ Evolution of the phase fluctuation $\delta y$ of the 
    complex scalar field 
    with initial deviation $\Delta=0$.
  \label{fig-com-fl-Y} } 
\end{figure}

\begin{figure}
    \caption{ Power spectrum of the radial fluctuation 
    $\delta x_k$ for (a) $a \simeq 1000$, (b) $a \simeq 2000$, 
    (c) $a \simeq 3000$ and (d) $a \simeq 3500$. 
  \label{fig-com-sp} } 
\end{figure}

\begin{figure}
    \caption{ Power spectrum of the fluctuation $\delta y_k$ for 
    (a) $a \simeq 1000$, (b) $a \simeq 2000$, (c) $a \simeq 3000$ and
    (d) $a \simeq 3500$. 
  \label{fig-com-spY} } 
\end{figure}

\begin{figure}
    \caption{ Evolution of the effective mass square at $\Phi=0$
    for the complex scalar field 
    with initial deviation $\Delta=0$.
  \label{fig-com-mphi} }
\end{figure}

\begin{figure}
    \caption{ Critical value of $\Delta$ as a function of
    $\eta$. The stars are from numerical calculations. 
    The dashed line shows the analytic estimation without fluctuations 
    (see Fig.~\ref{fig-scaling}),
    and the dotted line denotes the amplitude of the initial
    fluctuations produced in the inflationary epoch. 
  \label{fig-complex-scale1} }
\end{figure}

\begin{figure}
    \caption{ Parameter space for formation/ no formation of
    topological defects in the $g-\eta$ plane. NA denotes the region
    where we do not have the result, and OC is the region where the
    initial value of the field is less than the breaking scale: 
    $|\Phi(0)| < \eta$. In the OC region, the field already settles
    down to one the minimum of the potential at the end of inflation.
  \label{fig-complex-scale2} } 
\end{figure}

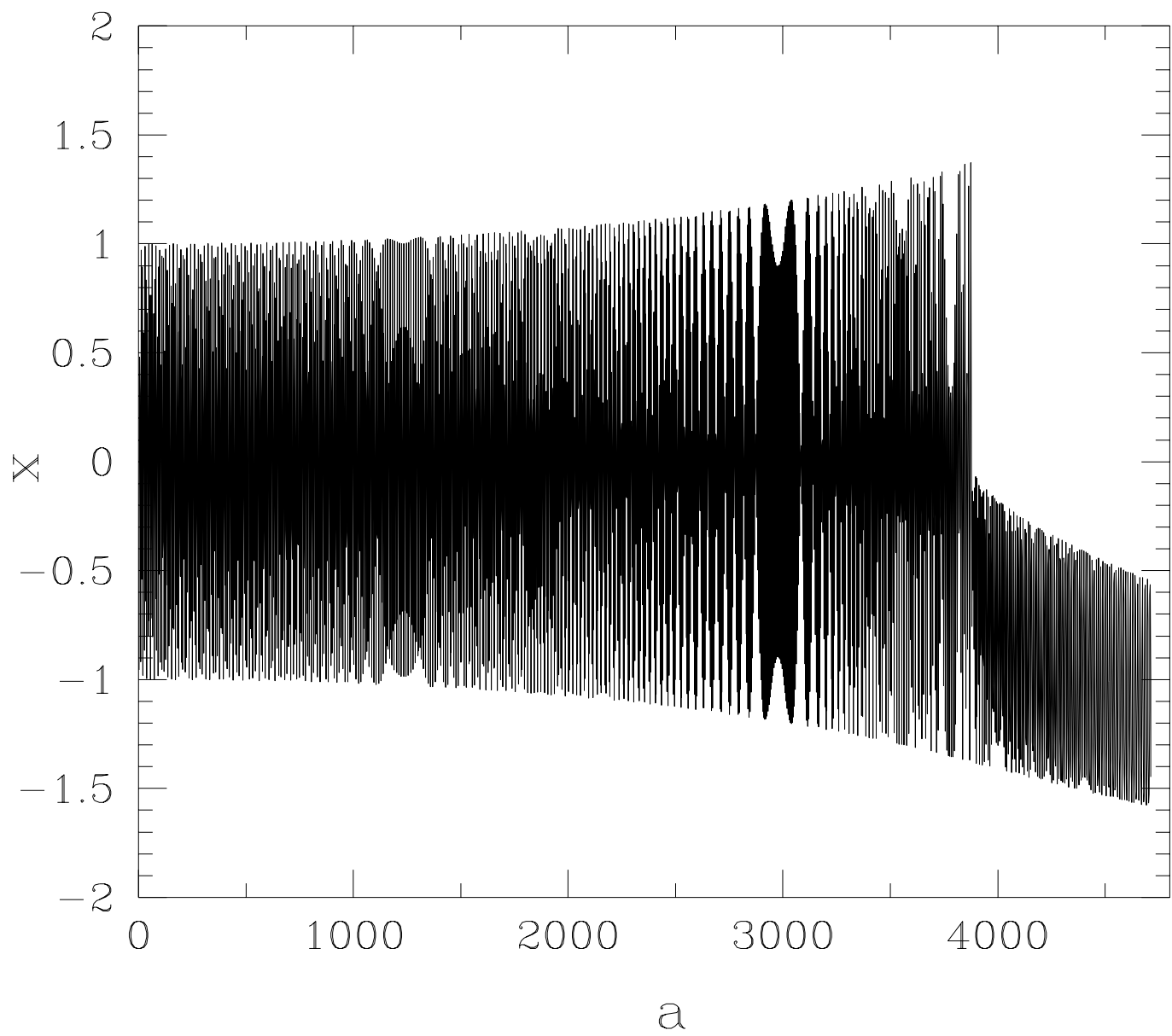
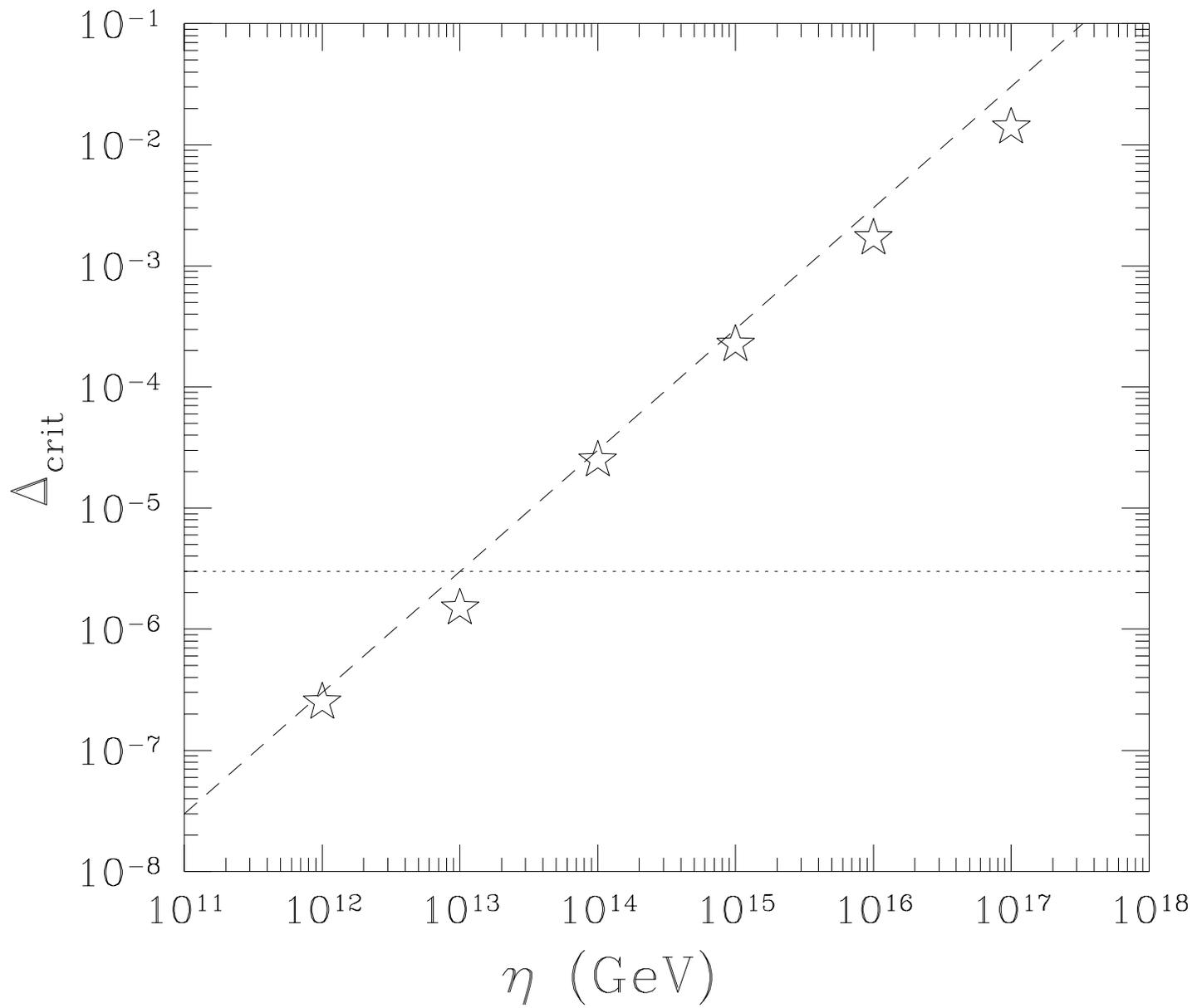
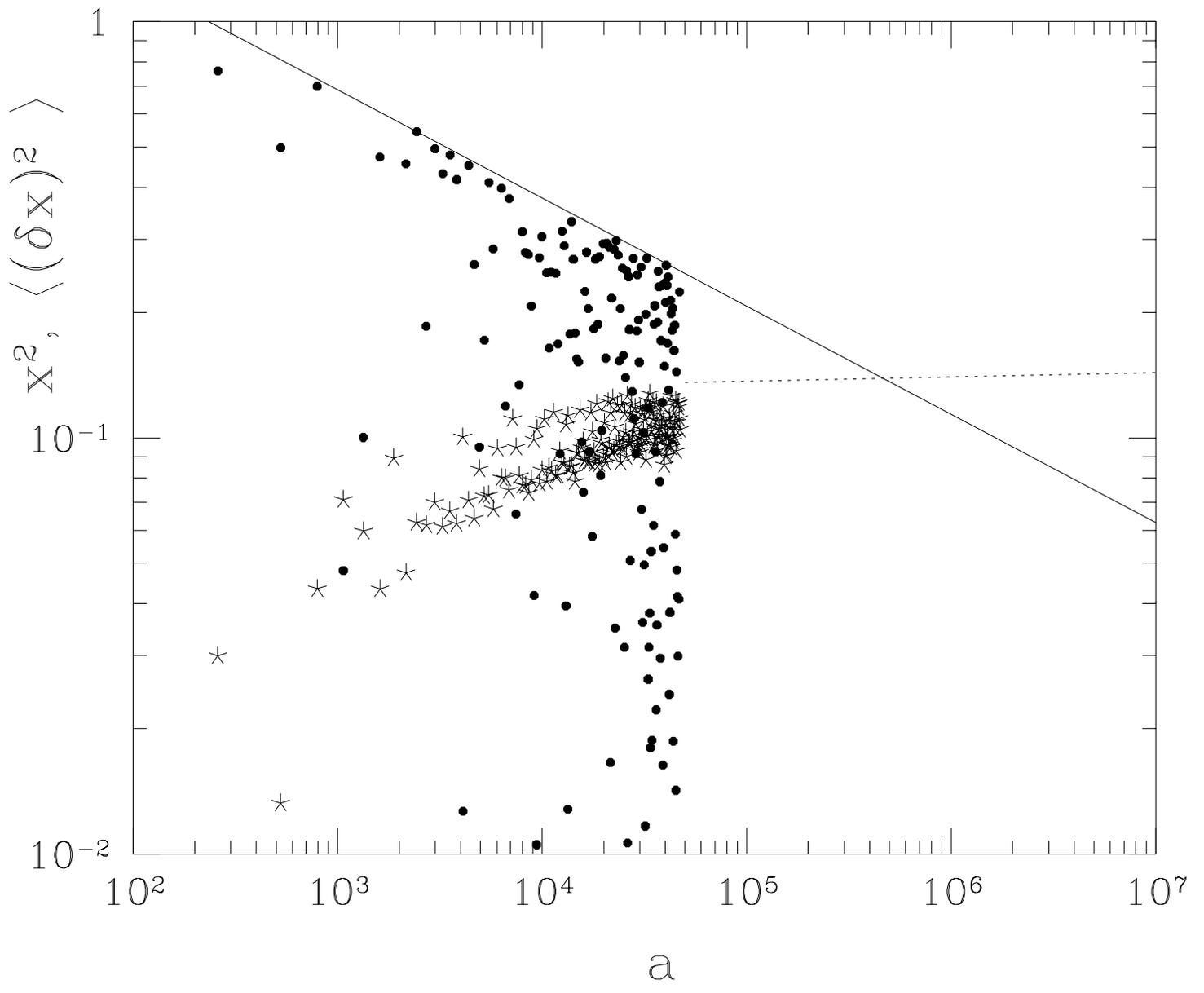
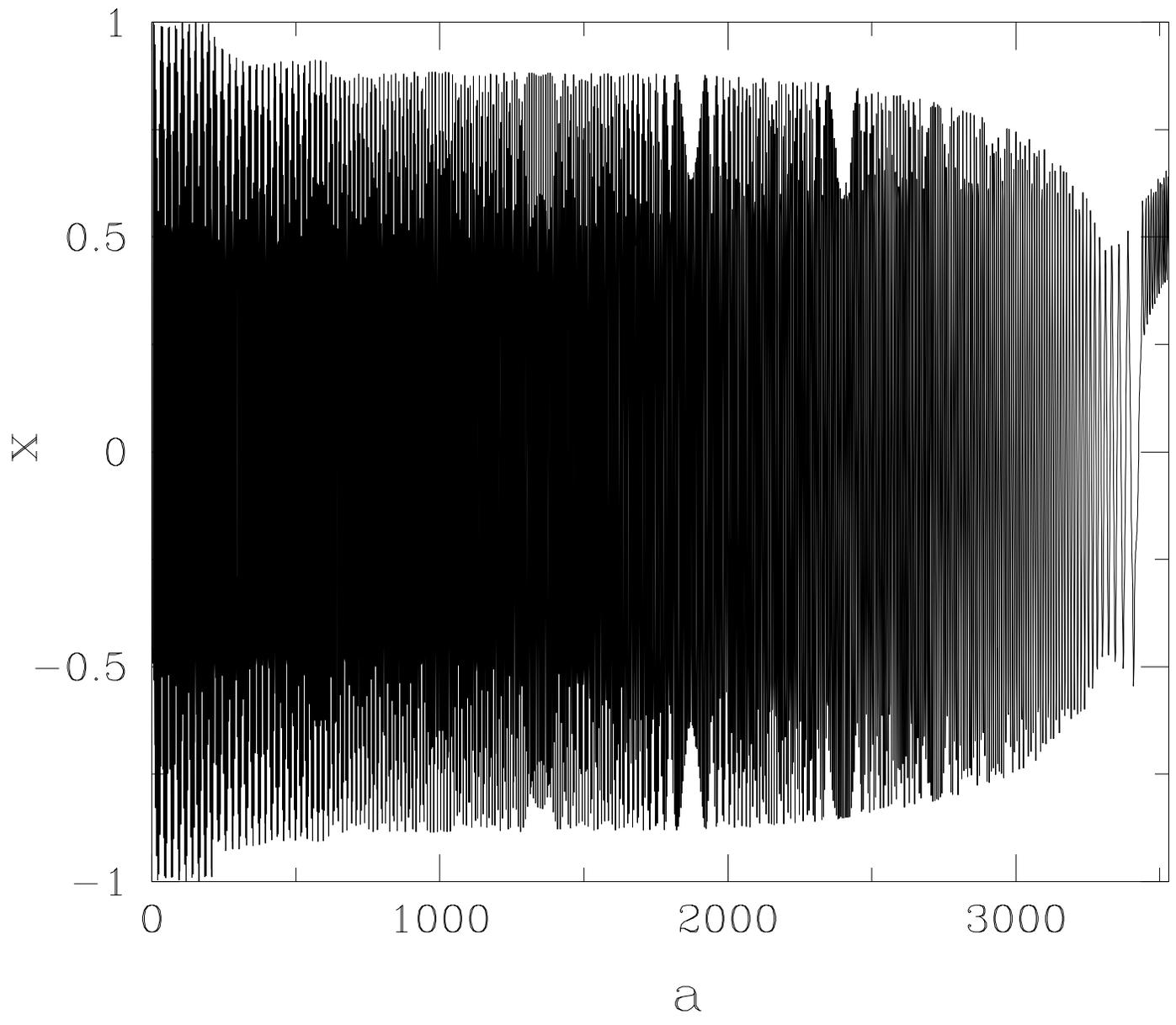
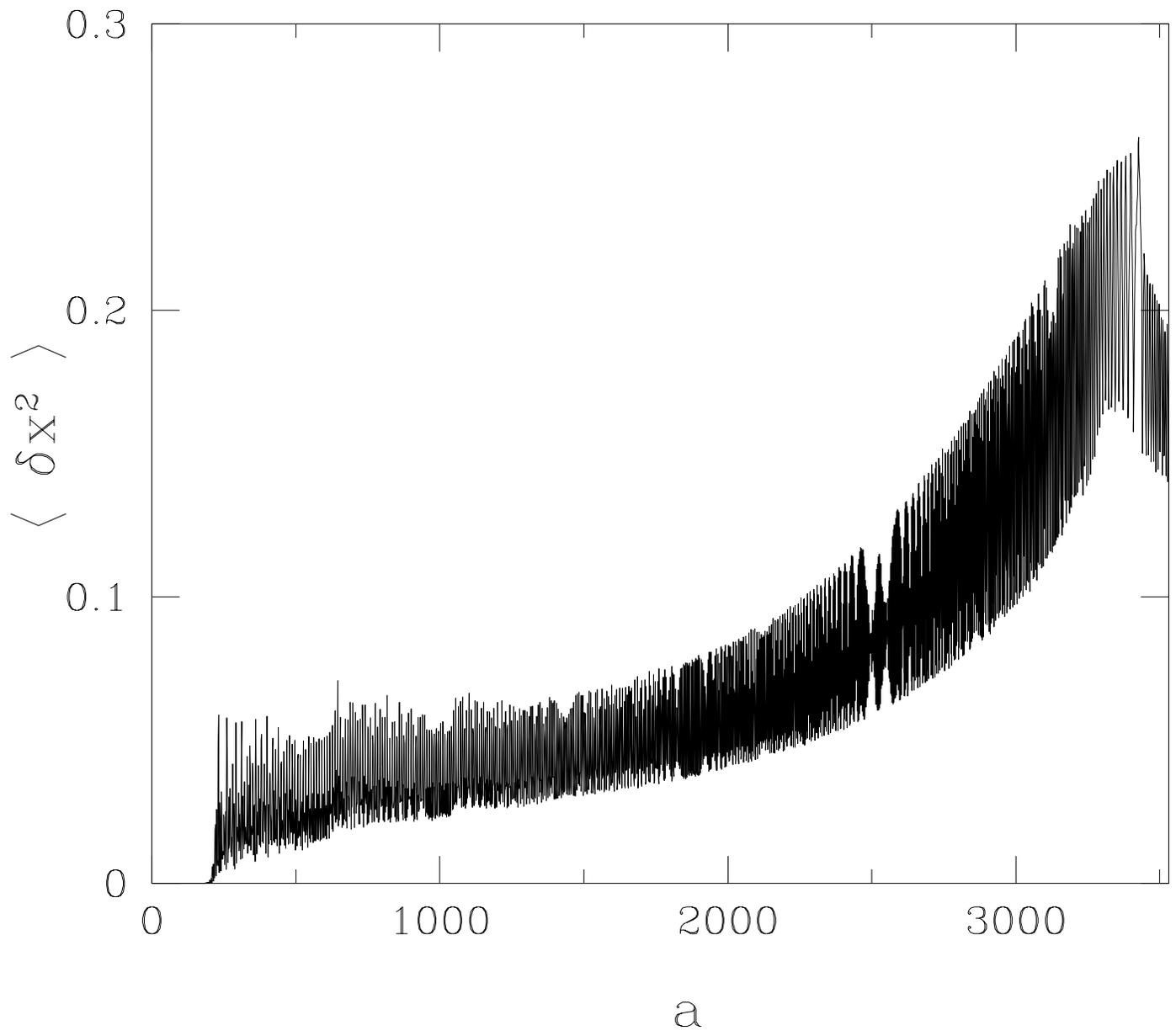
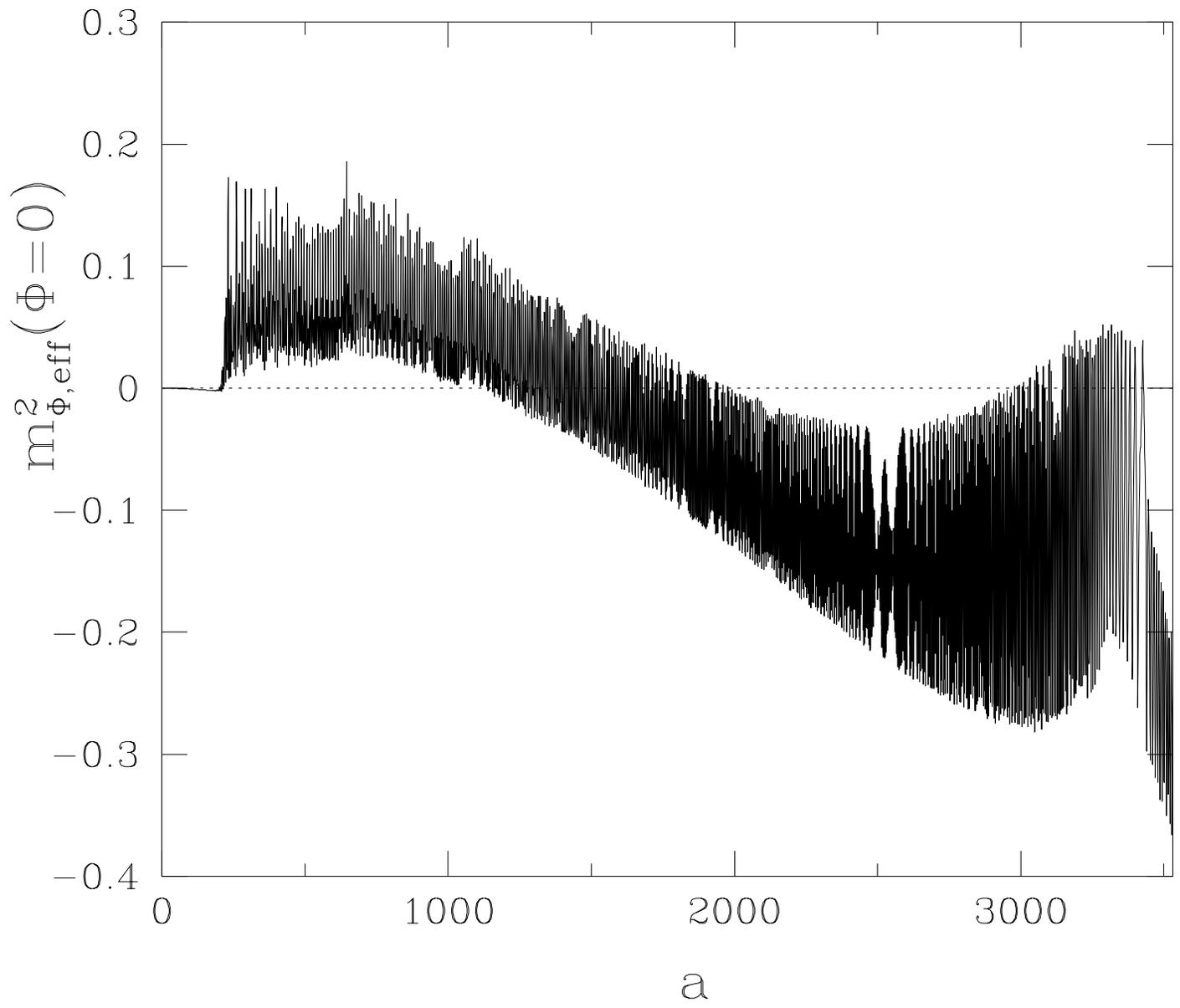
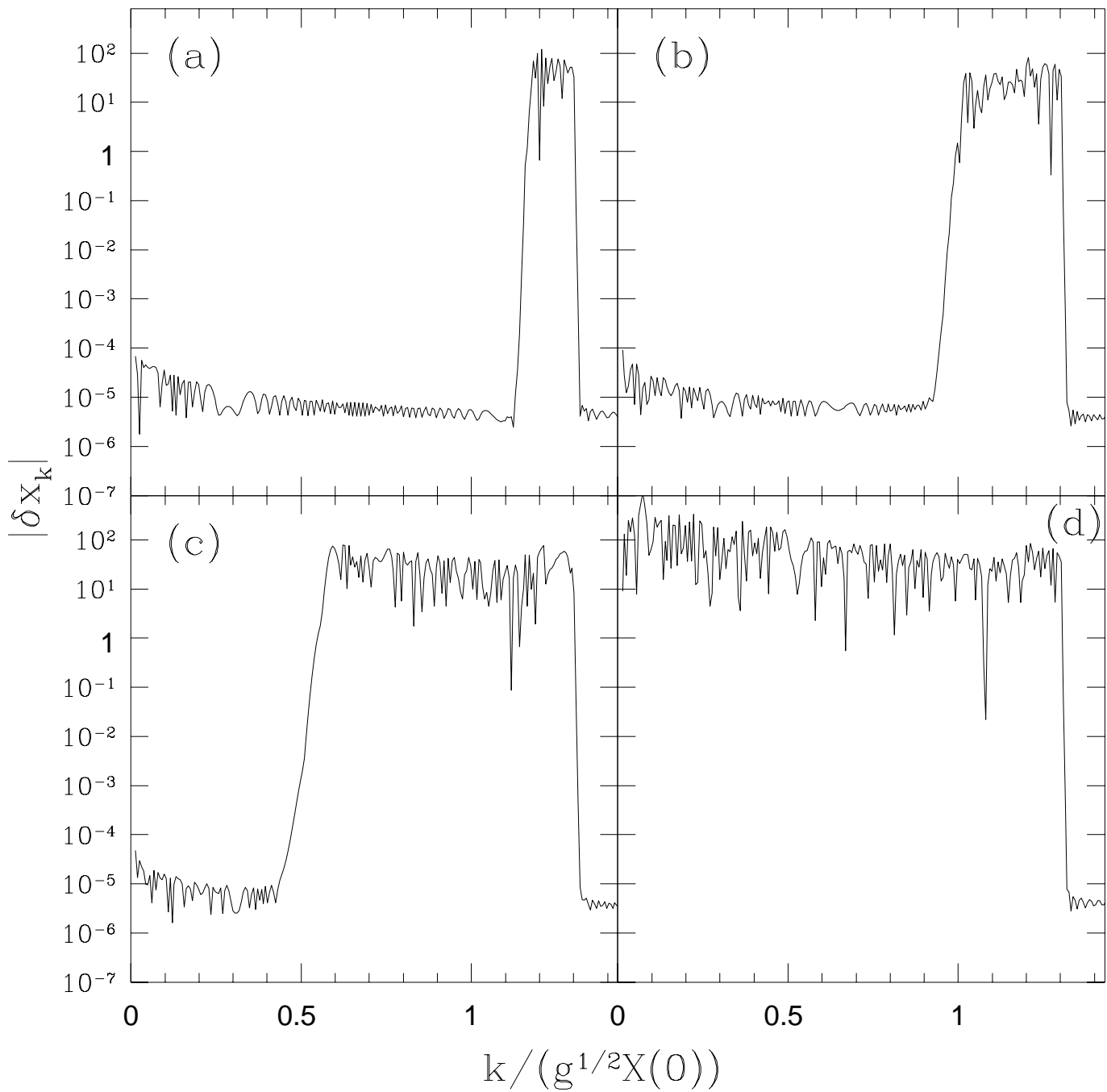
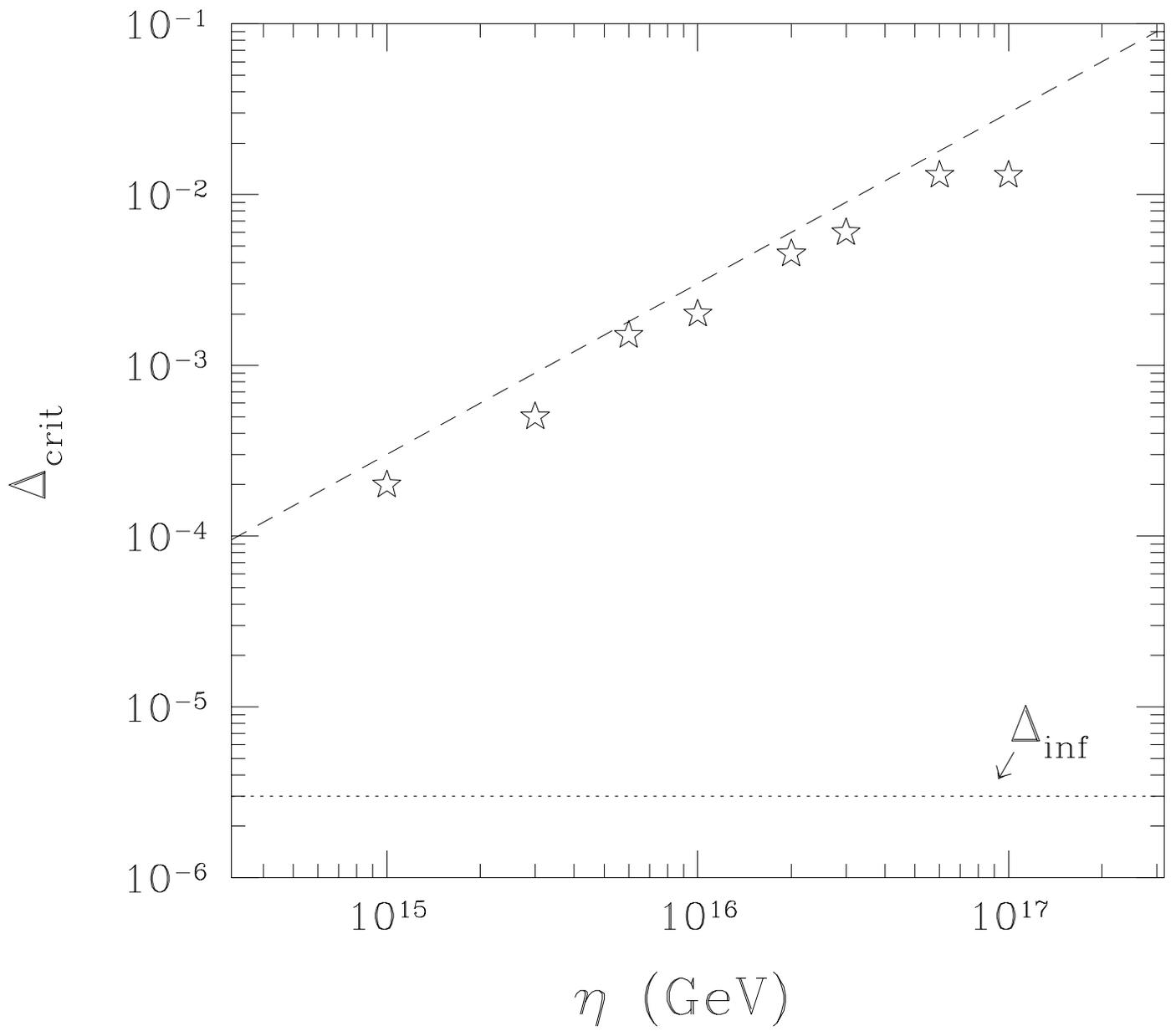
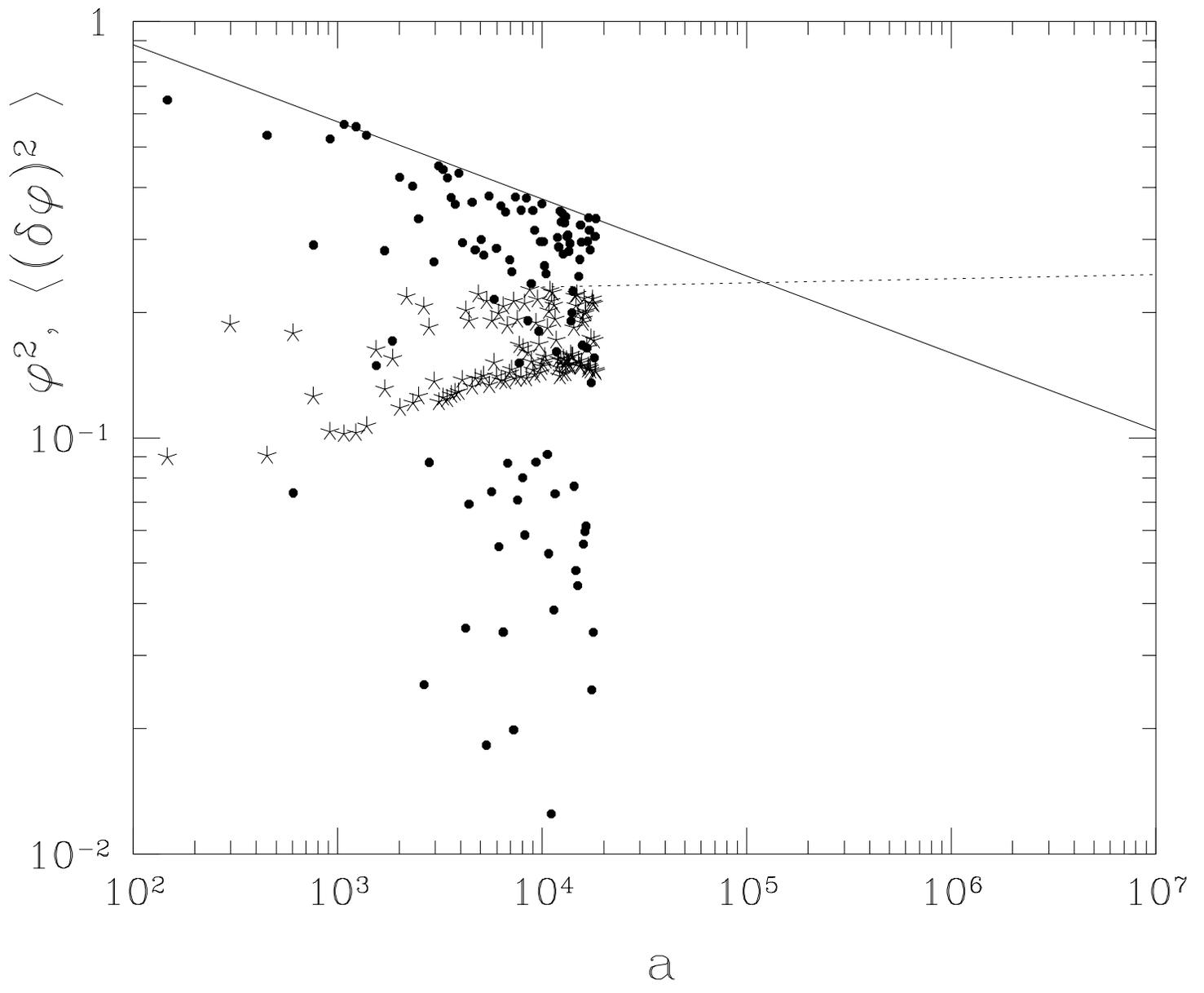
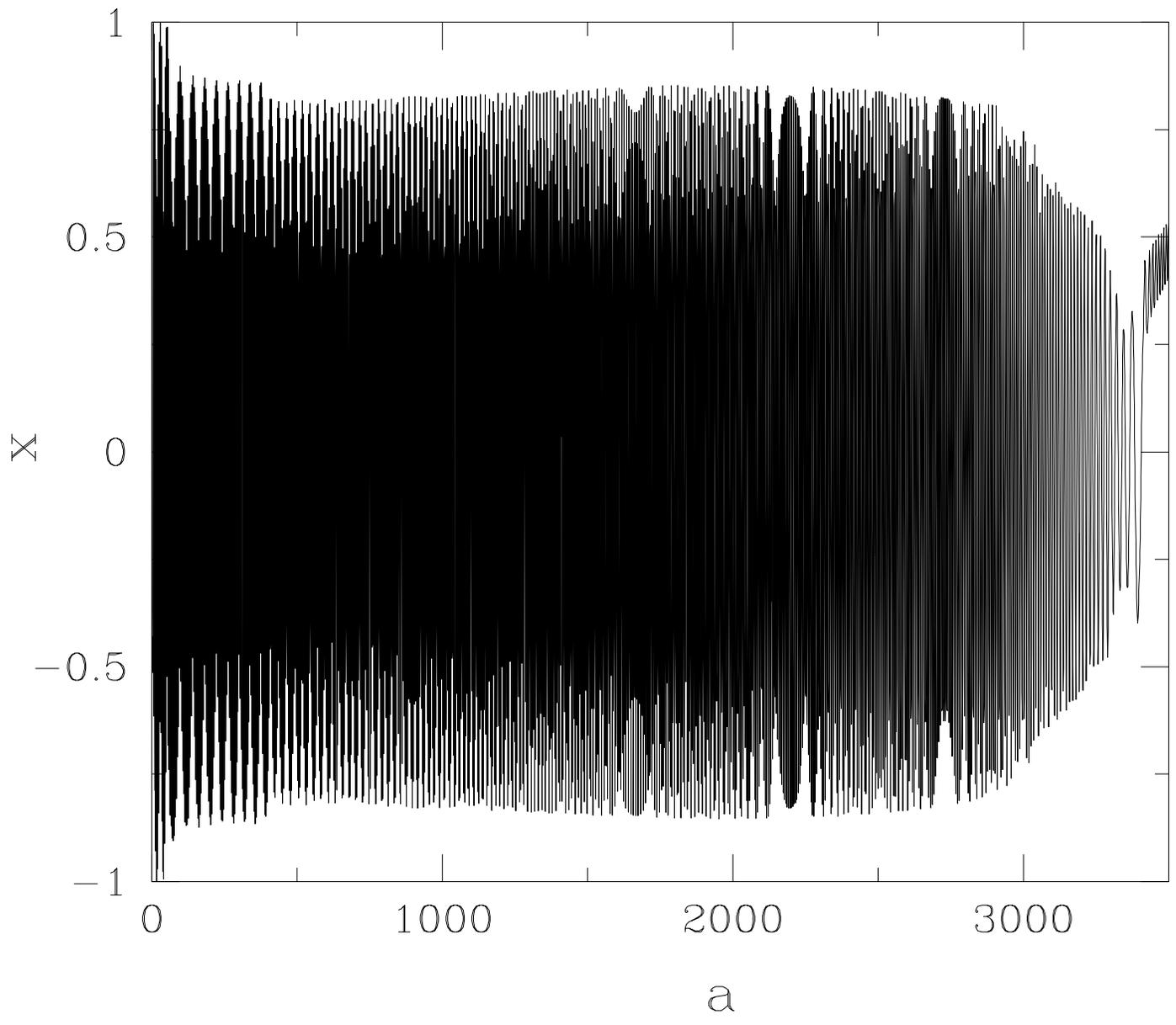
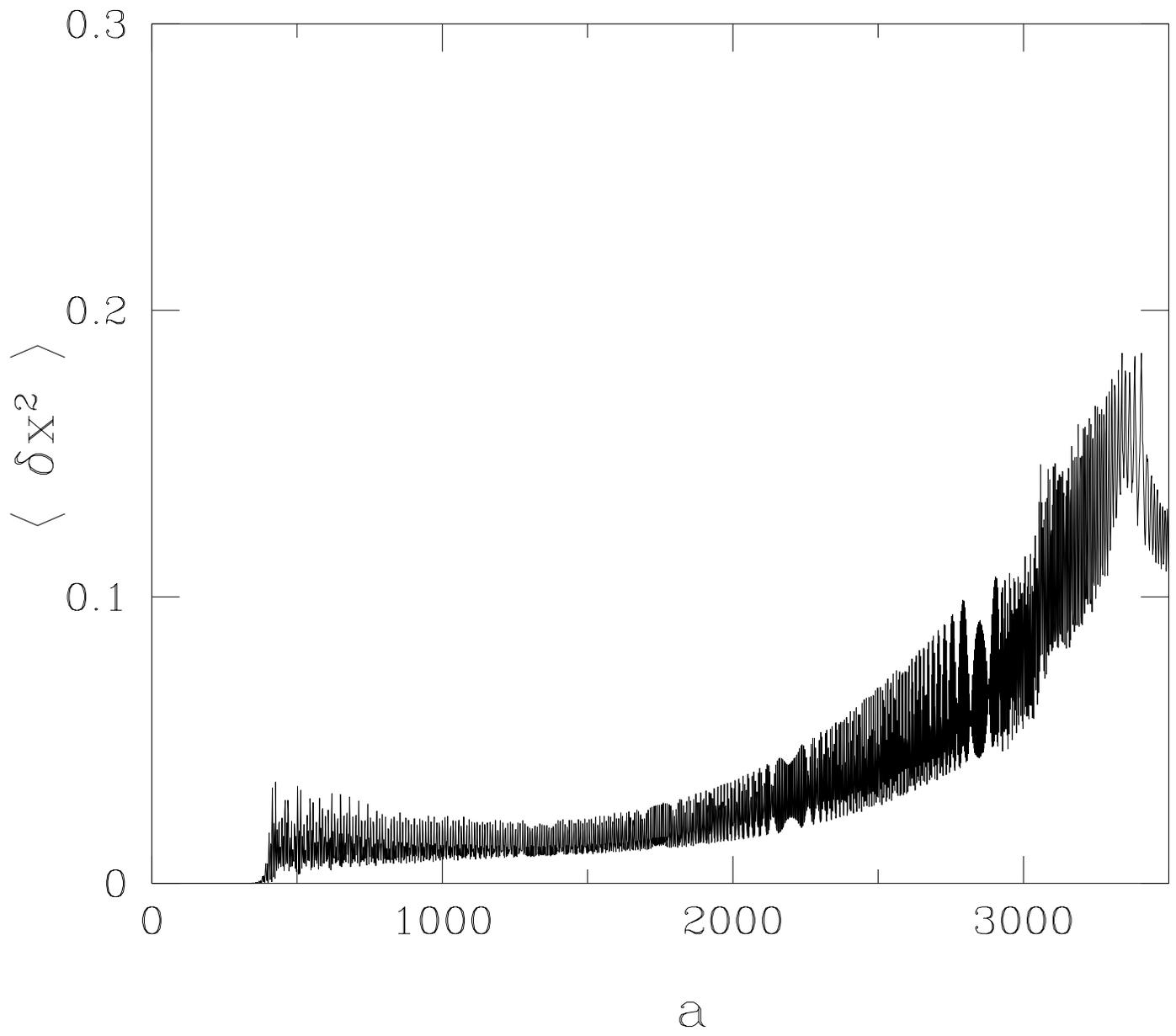
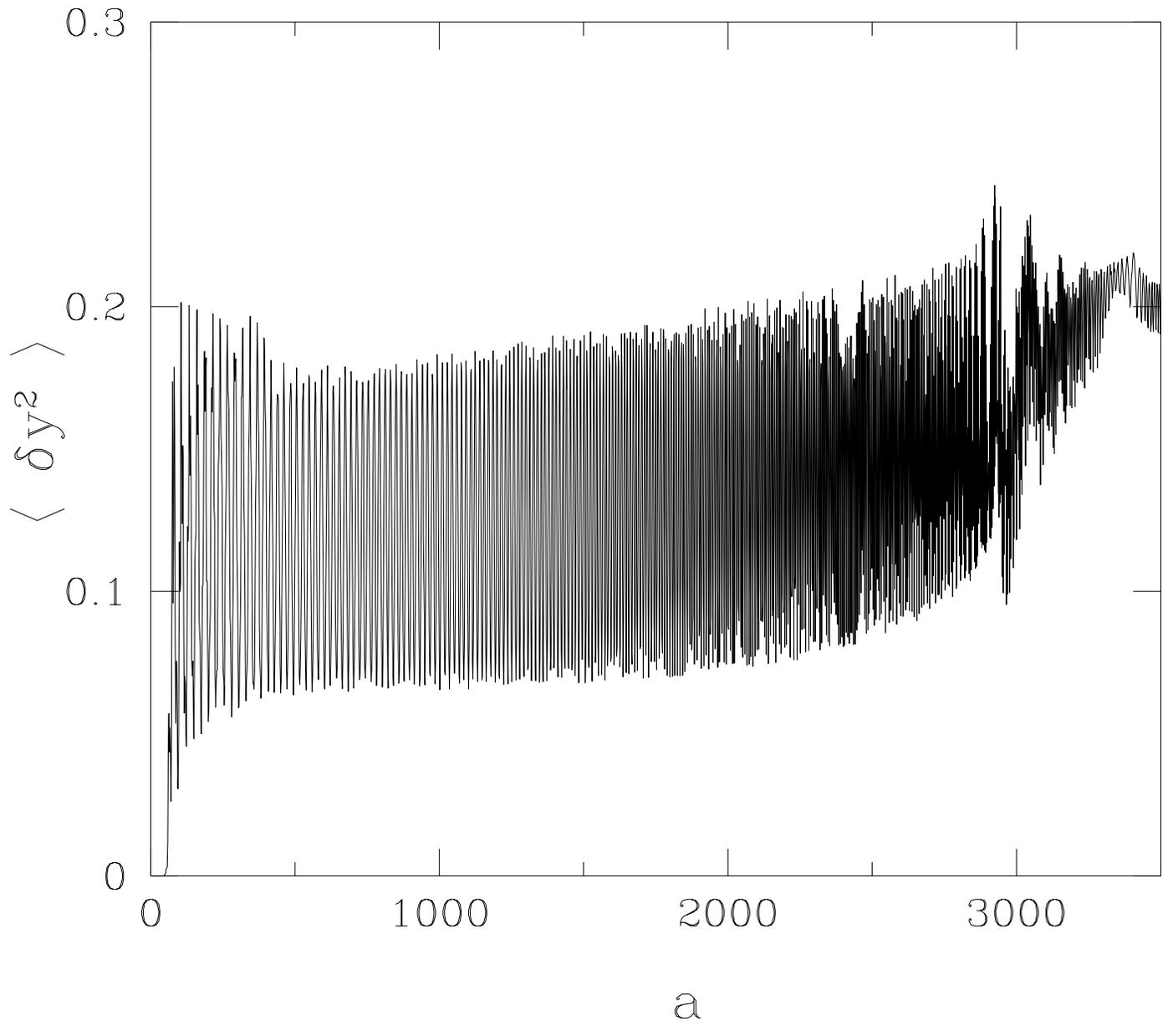
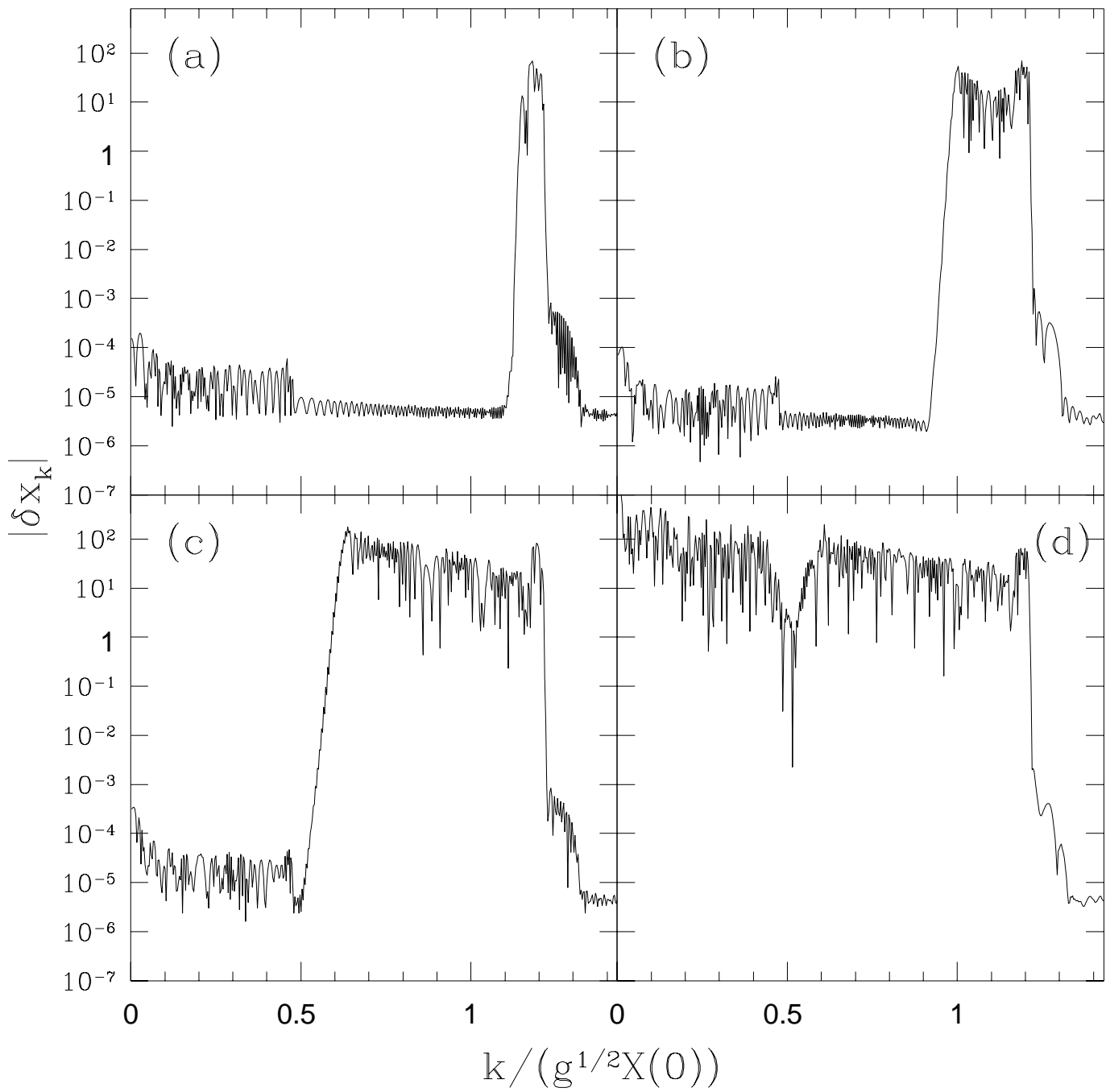
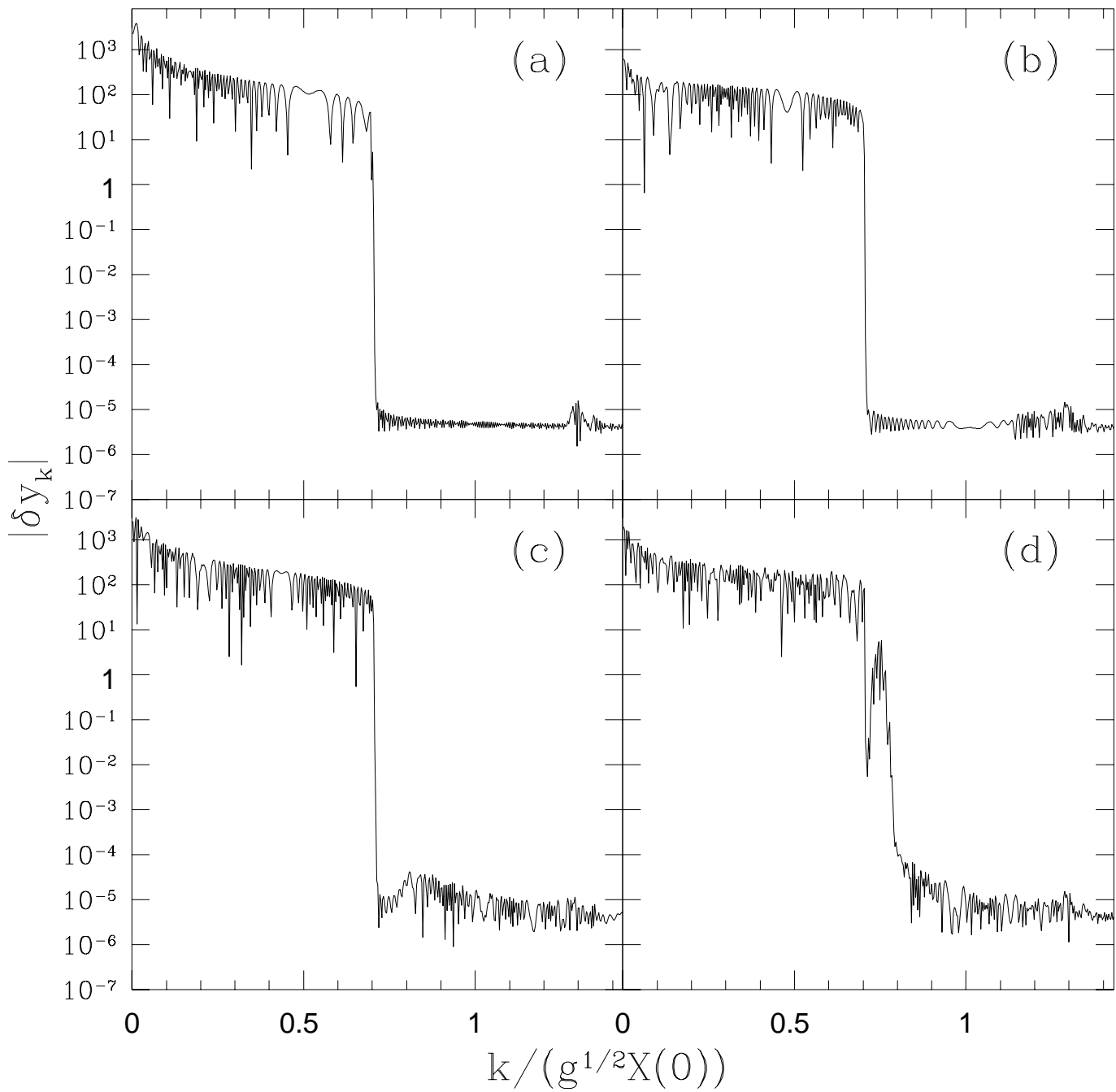
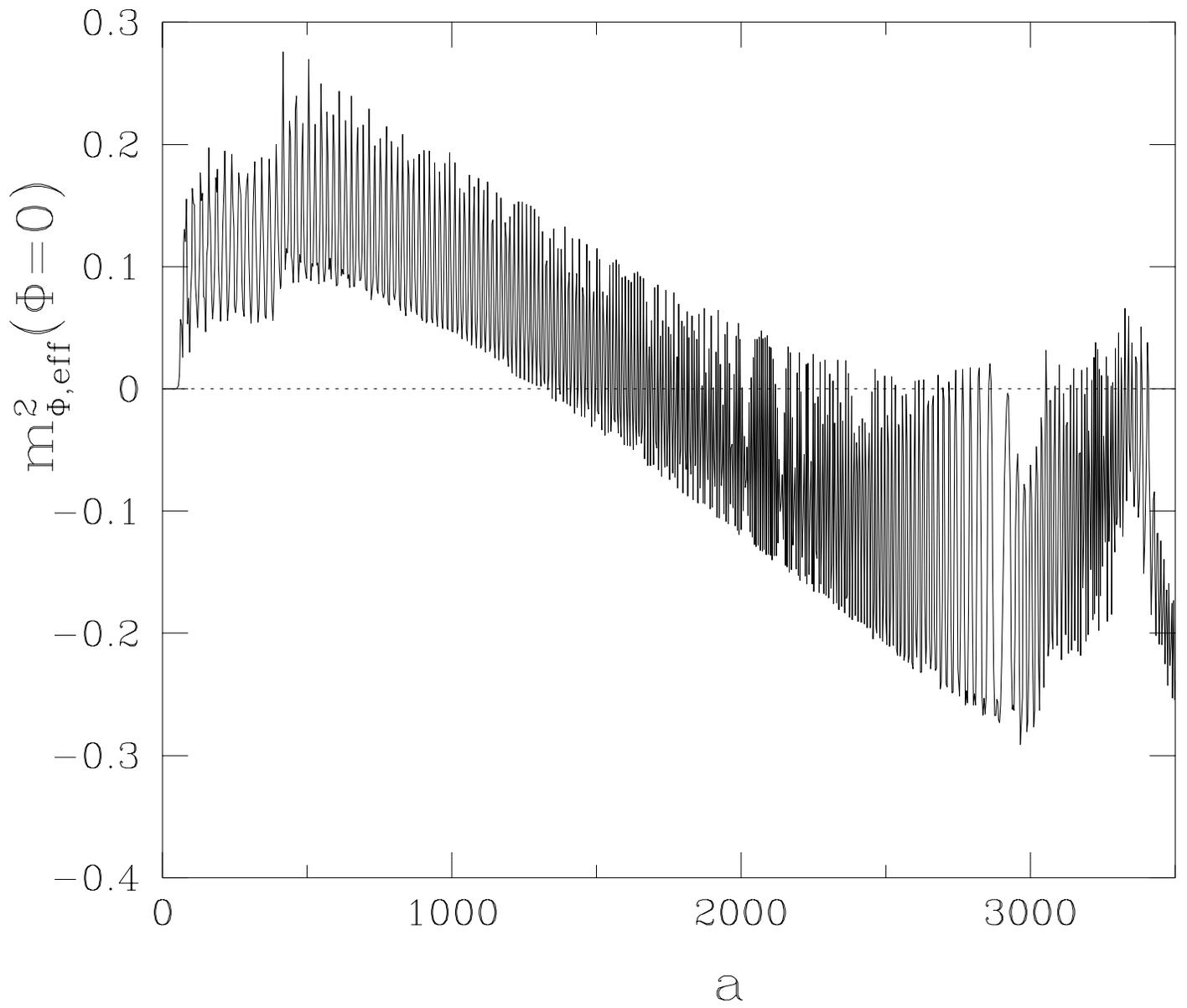
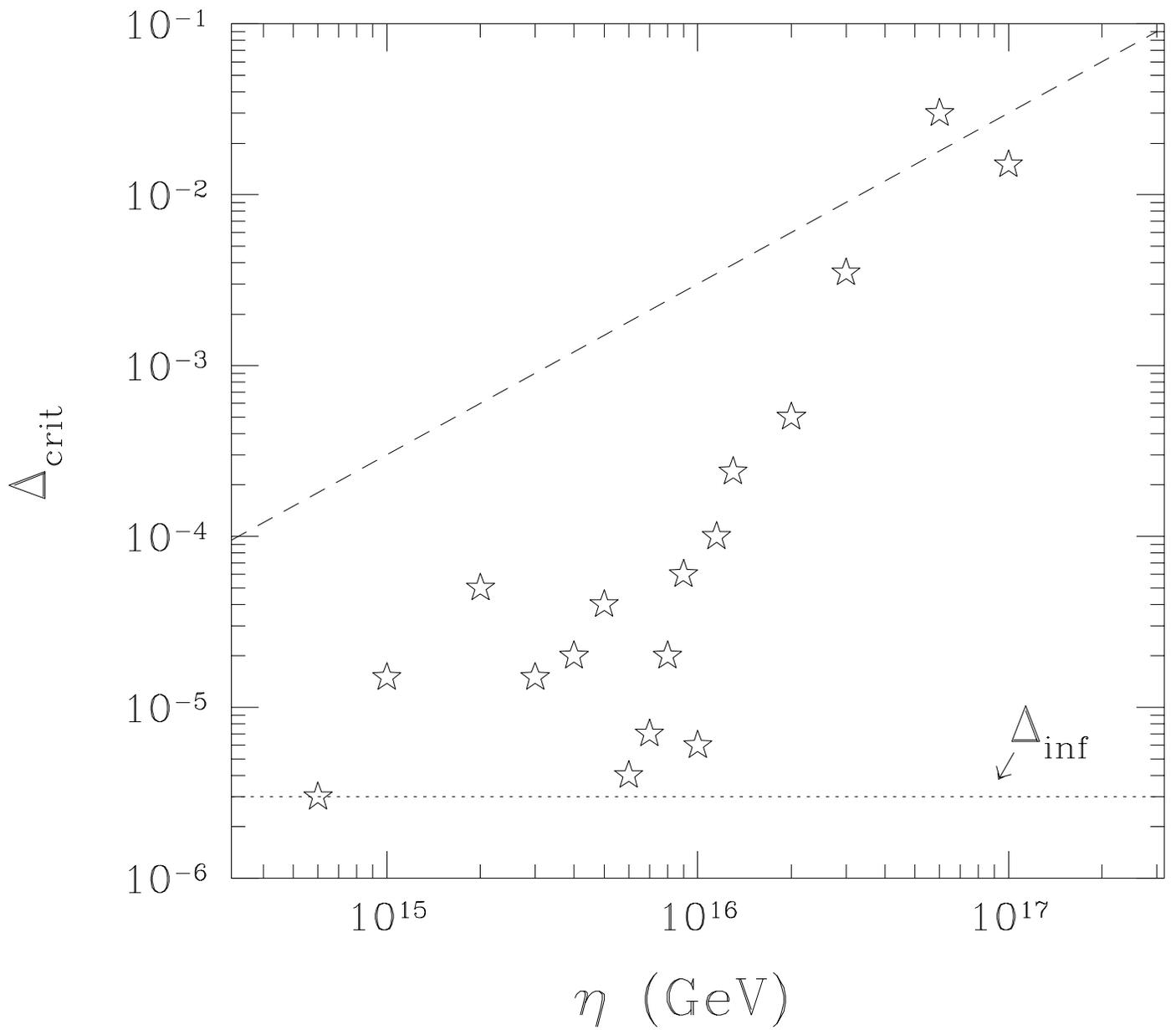
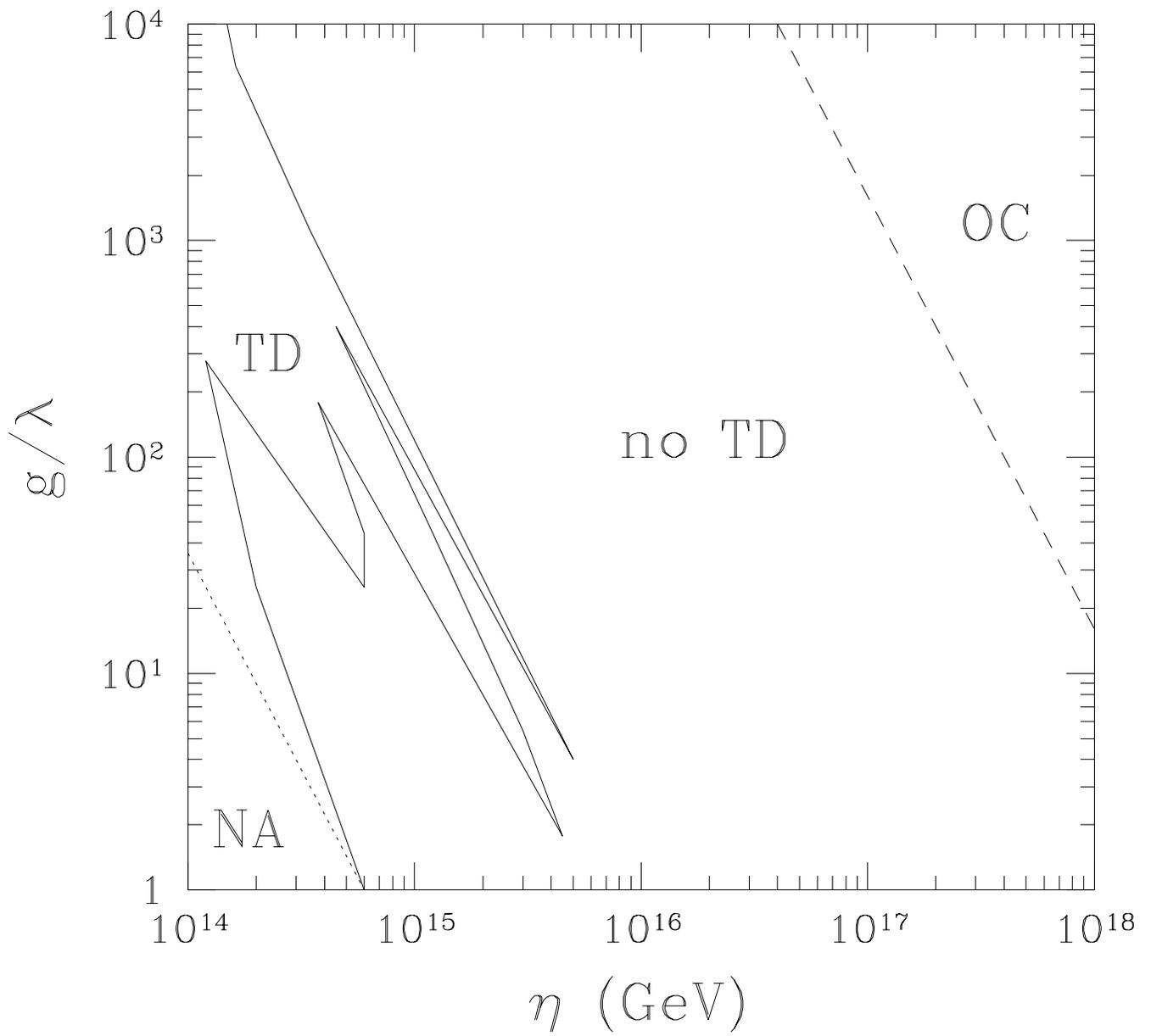
\end{document}